\documentclass[twocolumn,amsmath,amssymb,nofootinbib,10pt]{revtex4-1}
\pdfoutput=1
\usepackage{graphicx,float,amsmath,amsmath, amssymb}
\usepackage[braket]{qcircuit}
\usepackage[export]{adjustbox}

\def\be{\begin{equation}}
\def\ee{\end{equation}}
\def\bea{\begin{eqnarray}}
\def\eea{\end{eqnarray}}
\def\bse{\begin{subequations}}
\def\ese{\end{subequations}}

\usepackage[breaklinks, colorlinks,linkcolor=black, citecolor=black,urlcolor = black]{hyperref}
\usepackage[labelsep=period]{caption}
\usepackage{subcaption}
\usepackage{wasysym}
\usepackage{multirow}
\usepackage{soul}

\graphicspath{{pictures/}}
\begin{document}
\title{Quantum simulations of a qubit of space}

\author{Grzegorz Czelusta}%
\author{Jakub Mielczarek}%
\email{jakub.mielczarek@uj.edu.pl}
\affiliation{
Institute of Theoretical Physics, Jagiellonian University, {\L}ojasiewicza 11, 30-348 Cracow, Poland}

\date{\today}
\begin{abstract} 
In loop quantum gravity approach to Planck scale physics, quantum geometry is 
represented by superposition of the so-called spin network states. In the recent 
literature, a class of spin networks promising from the perspective of quantum 
simulations of quantum gravitational systems has been studied. In this case, 
the spin network states are represented by graphs with four-valent nodes,
and two dimensional intertwiner Hilbert spaces (qubits of space) attached to 
them. In this article, construction of quantum circuits for a general intertwiner 
qubit is presented. The obtained circuits are simulated on 5-qubit (Yorktown) 
and 15-qubit (Melbourne) IBM superconducting quantum computers, giving 
satisfactory fidelities. The circuits provide building blocks for quantum simulations 
of complex spin networks in the future. Furthermore, a class of maximally 
entangled states of spin networks is introduced. As an example of application, 
attempts to determine transition amplitudes for a monopole and a dipole spin 
networks with the use of superconducting quantum processor are made. 
\end{abstract}
\maketitle

\section{Introduction}

In the recent articles \cite{Li:2017gvt,Mielczarek:2018ttq,Mielczarek:2018nnd,
Mielczarek:2018jsh,Cohen:2020jlj,Zhang:2020lwi} an idea of performing quantum 
simulations of loop quantum gravity (LQG) \cite{Ashtekar:2004eh,Rovelli:1997yv} 
has been developed. While at present such simulations are possible to execute only for 
very simple systems, the approach may provide a way to investigate collective 
properties of Planck scale degrees of freedom in the future. 

Taking into account exponential growth of the dimensionality of the Hilbert space 
with the increase of the involved degrees of freedom, simulation of complex quantum 
gravitational systems is an extremely difficult task for classical computers. On the 
other hand, the current progress in quantum computing technologies may open a way
to simulate quantum gravitational systems unachievable to the most powerful classical 
supercomputers yet in this decade. Such claim is supported by the recent results of
quantum computations of the sampling problem from a quasi-random quantum circuit 
performed on a 53 qubit quantum processor \cite{GoogleSup}. Therefore, even if available 
quantum computing resources are still very limited, it is justified to already now prepare, 
test and optimize quantum circuits for the future quantum simulations of the Planck scale 
physics. A side benefit of such investigations is exploration of the quantum information 
structure of geometry, within and beyond LQG. In particular, the studies may shed a 
new light on such fundamental issues as emerging Gravity/Entanglement duality 
\cite{Ryu:2006bv,Swingle:2009bg} and related ER=EPR conjecture \cite{Susskind:2017ney}. 
The duality has its roots in the holographic principle \cite{Susskind:1994vu} 
and AdS/CFT correspondence \cite{Maldacena:1997re}.

Following the correspondence's holographic nature, the gravitational 3D bulk geometry is 
dually described by the 2D boundary. From this point of view, the LQG spin networks can 
be interpreted as the representations of either a state of gravity in bulk or, equivalently, the 
entanglement structure (similarly to tensor networks) of the boundary \cite{Han:2016xmb}. 
In consequence, simulating quantum gravity on a quantum computer may concern either the 
3D bulk or the 2D boundary. In the latter case, simulations of a quantum system at the 
boundary (e.g., a specific spin system) should allow reconstructing a state of quantum 
geometry in the bulk. It is, therefore, worth emphasizing that quantum simulations of 2D 
gravitational surfaces are of particular interest. The first attempt at quantum simulations 
in the holographic context has already been made in Ref. \cite{Li:2017qwu}. Concerning 
LQG, an example of a relevant quantum model of a boundary has been introduced in Ref. 
\cite{Feller:2017ejs}. The model utilizes the intertwiner degrees of freedom investigated 
here. Our studies may, therefore, be considered as a vestibule to quantum simulations of 
this and other similar models of quantum boundaries (but also the bulk geometry) in the 
future.

\begin{figure}[ht!]
\centering
\includegraphics[scale=0.3]{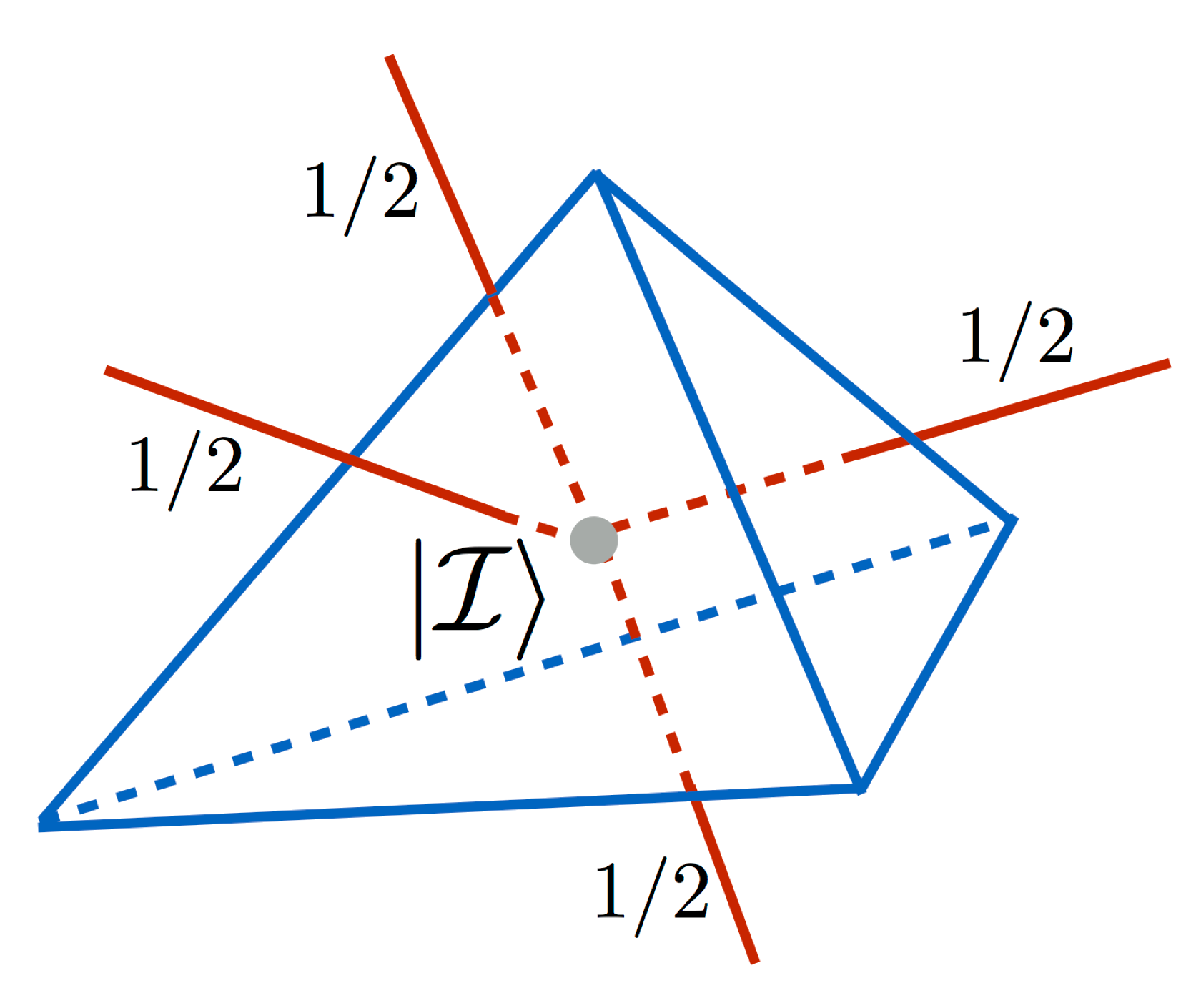}
\caption{4-valent node of the spin network and its geometrical interpretation as a tetrahedon.}
\label{Node}
\end{figure}

In this article, we follow the discussion presented in \cite{Mielczarek:2018jsh} where 
a class of spin networks characterized by 4-valent nodes has been considered. 
It has been shown, that while spin labels at the links are given by fundamental 
representations of the SU(2) group, the intertwiner spaces at the nodes are two 
dimensional Hilbert spaces. The Hilbert spaces are invariant subspaces (singlets) 
of four spin-1/2 Hilbert spaces associated with holonomies, which meet at the node
(see Fig. \ref{Node}). 

The singlet states are a consequence of the the local SU(2) gauge invariance 
imposed by the Gauss constraint in LQG, which has a form of a vector equation 
defining a tetrahedron (see e.g. \cite{CRFV}):
\begin{equation}
\sum_{i=1}^4\vec{J}_i =0, 
\label{GaussDef}
\end{equation}
where $\vec{J}_i$ are the angular momentum vectors normal to the faces of the tetrahedron. 
The components of the vector $\vec{J}_i=(J_i^1,J_i^2,J_i^3)$ satisfy the $\mathfrak{su}(2)$ 
algebra: $\{J_k^a,J_l^b \} =\delta_{kl} {\epsilon^{ab}}_{ c}J_k^c$, for $a,b,c \in \{1,2,3\}$.
The $\vec{J}_i$ vectors are associated to the areas $A_i := 8\pi l^2_\text{Pl} \gamma 
\sqrt{\vec{J}_i\cdot \vec{J}_i }$ of the faces, where $l_\text{Pl}:=\sqrt{G}\approx 1.62\cdot 10^{-35}$ m 
is the Planck length, $G$ is the Newton's constant and $\hslash=1=c$.
Here, $\gamma$ is the Barbero-Immirzi parameter \cite{Immirzi:1996dr,Barbero:1994ap}, 
which plays an important role in LQG\footnote{The Barbero-Immirzi parameter enters considerations 
via the Holst term in the gravitational action. The term is typically not-contributing (on-shell) to the
classical considerations. However, an exception is a case with the fermionic matter, in which the 
Holst term leads to (potentially observable) violation of parity \cite{Perez:2005pm,Freidel:2005sn}. 
Based on black hole entropy considerations in LQG, the value of $\gamma$ of the order of 
unity is expected \cite{Ashtekar:1997yu,Meissner:2004ju,Agullo:2010zz}.}.

In such a case, a general intertwiner state - an intertwiner qubit - can be written as \cite{Mielczarek:2018jsh}:  
\begin{equation}
| \mathcal{I} \rangle = \cos(\theta/2) |0_s\rangle+e^{i\phi} \sin(\theta/2)|1_s\rangle, \label{IntertwinerState}
\end{equation} 
where $\theta \in [0,\pi]$ and $\phi \in [0, 2\pi)$ are angles on the Bloch sphere. 
The $|0_s\rangle$ and $|1_s\rangle$ are basis states, corresponding to two linearly 
independent singlets of four spin-1/2 DOFs (qubits) in the $s$-channel \cite{Feller:2015yta}:
\begin{align}
|0_s\rangle&= |S\rangle  |S\rangle, \label{0s} \\
|1_s\rangle&= \frac{1}{\sqrt{3}} \left( |T_{+}\rangle |T_{-}\rangle+
|T_{-}\rangle |T_{+}\rangle-|T_{0}\rangle |T_{0}\rangle   \right), \label{1s}
\end{align}
where 
\begin{align}
|S\rangle &= \frac{1}{\sqrt{2}} \left(|01\rangle-|10\rangle \right), \\
|T_{+}\rangle &= |00\rangle, \\
|T_{0}\rangle &= \frac{1}{\sqrt{2}} \left(|01\rangle+|10\rangle \right), \\
|T_{-}\rangle &= |11\rangle,
\end{align}
are two spin-1/2 singlet and triplet states respectively. The Hilbert space of the 
spin-1/2 DOF is $\mathcal{H}_{1/2} = \text{span} \{|0\rangle,|1\rangle  \}$. 

Physically, the intertwiner space is associated with the quantum of volume 
\cite{Rovelli:1994ge,Ashtekar:1997fb}. This can be shown by considering 
volume operator $\hat{V}$ in LQG, defined as follows \cite{CRFV}:
\begin{equation}
\hat{V}:= \frac{\sqrt{2}}{3} l_\text{Pl}^3  (8 \pi \gamma)^{\frac{3}{2}} \sqrt{ \left| \hat{\vec{J}}_1
\cdot ( \hat{\vec{J}}_2 \times  \hat{\vec{J}}_3) \right|},
\label{volumeoperator}
\end{equation}
where $\hat{\vec{J}}_i$ are the angular momentum vector operators. The volume 
operator is defined as a positive-definite function of the triple product $\vec{J}_1
\cdot (\vec{J}_2 \times \vec{J}_3)$, the sign of which depends on the orientation 
of space. The two possible signs discriminate between the two eigenvalues of the 
operator  $\hat{\vec{J}}_1\cdot ( \hat{\vec{J}}_2 \times  \hat{\vec{J}}_3)$. In order to 
keep this information at the level of the volume positive-definite operator $\hat{V}$, 
one can extend its definition (\ref{volumeoperator}) to the \emph{oriented volume} 
case. In consequence, the two signs will distinguish the two (initially degenerated)      
eigenvalues of the volume operator. One can find that, the following 
superpositions of the basis states $|0_s\rangle$ and $|1_s\rangle$:
\begin{eqnarray}
|V_+\rangle &=& \frac{1}{\sqrt{2}}\left(| 0_s \rangle - i | 1_s \rangle  \right), \label{Vplus} \\ 
|V_-\rangle &=& \frac{1}{\sqrt{2}}\left(| 0_s \rangle + i | 1_s \rangle  \right), \label{Vminus}
\end{eqnarray}
are eigenstates of the volume operator, such that the \emph{oriented} eigenvalues 
satisfy: $\hat{V}|V_+\rangle= +V_0|V_+\rangle$ and $\hat{V}|V_-\rangle= -V_0|V_-\rangle$  
\cite{CRFV}. The $V_0 := \frac{l^3_{\text{Pl}}(8 \pi \gamma)^{\frac{3}{2}}}{\sqrt{6 \sqrt{3}}}$ 
is a quantum of 3-volume in LQG. This justifies why we call the two dimensional intertwiner 
a \emph{qubit of space}.

Worth mentioning is that the intertwiner states are relevant in 
quantum information theory. Namely, encoding one logical qubit (the intertwiner qubit) 
in four physical qubits allows for quantum communication without a shared reference 
frame \cite{Bartlett}. Let $\rho_A$ be a state (i.e. density matrix) that Alice wants to 
send to Bob. Bob because of his lack of knowledge about Alice's reference 
frame receives state $\rho_B$:
\begin{equation}
	\rho_B=\int_GdgU\left(g\right)\rho_AU^\dagger\left(g\right),
\end{equation}
where $g\in G$ and $G$ is a group of transformations between the two reference frames, 
and $dg$ is the Haar measure. The operation $U\left(g\right) :=U_1\left(g\right)\otimes U_2\left(g\right) 
\otimes U_3\left(g\right)\otimes U_4\left(g\right)$ is a tensor product of (the same) single-qubit 
unitary operators $U_i(g)$ for $i=1,2,3,4$, acting on the four composite qubits of the intertwiner 
states.  In the case when Alice and Bob share no knowledge about the orientation 
of their frames, we have $G=SU\left(2\right)$. Consequently, one finds that in order to 
have $\rho_A=\rho_B$, the states invariant under the action of this group must be considered.
The is satisfied by the \emph{intertwiner qubits} considered here. Further discussion of this
property in the quantum gravitational context can be found in Ref. \cite{Livine:2005mw}.

In Ref. \cite{Mielczarek:2018jsh} a quantum circuit for the $|0_s\rangle$ basis state 
has been investigated and simulated on the IBM Q 5-qubit quantum processor. In this 
article, the analysis is extended to the general intertwiner qubit $| \mathcal{I} \rangle$ 
given by Eq. \ref{IntertwinerState}. In Sec. \ref{Sec:QuantumCircuit} a quantum circuit
for a general intertwiner state is introduced. Then, in Sec. \ref{Sec:Transpilation} the 
circuit is transpilated such that it fits to the topologies of the superconducting IBM 
quantum processors. The Sec. \ref{Sec:Examples} presents reduced forms of the 
quantum circuits for the special cases of the basis states: $|0_s\rangle$ and 
$|1_s\rangle$.  In Sec. \ref{Sec:Simulations} six representative states of the intertwiner
qubit are simulated on IBM 5 and 15 qubit quantum processors, which are available 
for cloud computing. Then, in Sec. \ref{Sec:Amplitudes} a general discussion of the 
transition amplitudes between the spin network states is given. A class of maximally 
entangled states which introduce quantum correlations between intertwiner qubits 
is introduced in Sec. \ref{Sec:EntangledNetwork}. The maximally entangled states 
are applied to the special cases of the monopole (Sec.\ref{Sec:Monopole}) and dipole 
(Sec. \ref{Sec:Dipole}) spin networks, for which attempts to determine transition 
amplitudes with the use of superconducting IBM quantum processors are made. 
Our results are summarized in Sec. \ref{Sec:Summary}. The article is accomplished 
with two appendices. Appendix A contains numerical results obtained from simulations 
of the interwiner qubits states, discussed in Sec. \ref{Sec:Simulations}. In Appendix B, 
results of test performed on the 15-qubit IBM quantum computer Melbourne are shown.  

\section{Quantum circuit} 
\label{Sec:QuantumCircuit}

The purpose of this section is to find quantum circuit representation of the unitary 
operator $\hat{U}_{\mathcal{I}}$, such that:
\begin{equation}
| \mathcal{I} \rangle = \hat{U}_{\mathcal{I}} |0000\rangle.  \label{stateprep} 
\end{equation}
Here, the  $| \mathcal{I} \rangle$ is a general intertwiner qubit state $| \mathcal{I} \rangle 
\in \mathcal{H}_{\mathcal{I}}:=\text{span}\{ | 0_s\rangle, | 1_s\rangle  \}$, given by 
Eq. \ref{IntertwinerState} and $|0000\rangle$ is the initial state of the quantum register.

The $\hat{U}_{\mathcal{I}}$ is a state preparation operator. The 
procedure of preparing $| \mathcal{I} \rangle $ is, however,  not unique since 
there are infinitely many operators $\hat{U}_{\mathcal{I}}$ that satisfy Eq. 
\ref{stateprep}. This is because only first column in the matrix representation 
of $\hat{U}_{\mathcal{I}}$ is fixed and there are still $n^2-2n-1=223$ undetermined 
free real parameters (here $n = \text{dim} \otimes \mathcal{H}_{1/2}^4 =16$ and 
the total irrelevant phase has also been subtracted). Furthermore, in general, 
expressing an operator in terms of quantum gates is a difficult task. Here, the 
goal is achieved by utilising some properties of the state $|\mathcal{I}\rangle$, 
which allows for systematic expressing of the state in terms of the elementary 
quantum gates acting on the initial state $|0000\rangle$.

Worth mentioning at this point is that, in general, one could expect that  some ancilla 
qubits may also be involved. However, as we will show here, additional logical qubits 
are not required to produce the state $| \mathcal{I} \rangle $. However, while noisy 
physical qubits are considered, quantum error correction codes need to be involved, 
which unavoidably utilize additional physical qubits. In this article, we will restrict our
considerations to the level of logical qubits and the quantum error correction codes 
will not be discussed.

In order to find the quantum circuit representing the operator $ \hat{U}_{\mathcal{I}}$
let us first apply Eqs. \ref{0s} and \ref{1s} to Eq. \ref{IntertwinerState}, 
which leads to: 
\begin{align}
| \mathcal{I} \rangle  &= \frac{c_1}{\sqrt{2}} ( |0011\rangle +|1100\rangle )  \nonumber \\ 
                                 &+ \frac{c_2}{\sqrt{2}} ( |0101\rangle +|1010\rangle ) \nonumber \\ 
                                 &+ \frac{c_3}{\sqrt{2}} ( |0110\rangle +|1001\rangle ), \label{Ipairs}
\end{align}
where coefficients $c_1$, $c_2$ and $c_3$ are complex-valued coefficients expressed as follows:
\begin{align}
c_1 &= \sqrt{\frac{2}{3}} e^{i\phi} \sin(\theta/2),   \label{c1}    \\
c_2 &=  \frac{1}{\sqrt{2}}\left( \cos(\theta/2)- \frac{1}{\sqrt{3}}e^{i\phi} \sin(\theta/2) \right) \nonumber \\
       &= \frac{e^{i\chi_{+}} }{\sqrt{2}}\sqrt{1-\frac{2}{3}\sin^2(\theta/2) -\frac{\sin \theta \cos \phi}{\sqrt{3}}},  \label{c2} \\
c_3 &=  \frac{1}{\sqrt{2}}\left( -\cos(\theta/2)- \frac{1}{\sqrt{3}}e^{i\phi} \sin(\theta/2) \right) \nonumber \\
       &= \frac{e^{i\chi_{-}}}{\sqrt{2}}\sqrt{1-\frac{2}{3}\sin^2(\theta/2)+\frac{\sin \theta \cos \phi}{\sqrt{3}}},  \label{c3}   
\end{align}
together with the phases 
\begin{align}
	\chi_{\pm} =&\arctan\left( \frac{\sin(\phi) \tan (\theta/2)}{\mp\sqrt{3}+\cos(\phi) \tan (\theta/2)}\right) \nonumber \\ 
	+&\frac{\pi}{2}\left[1-\text{sgn}\left(\pm\cos\left(\frac{\theta}{2}\right)
	-\frac{\cos\phi\sin\left(\frac{\theta}{2}\right)}{\sqrt{3}}\right)\right]. 
\end{align}
The coefficients (\ref{c1}), (\ref{c2}) and (\ref{c3}) satisfy the two conditions:
\begin{align}
\sum_{i=1}^3|c_i|^2=1, \ \text{and} \ \ \sum_{i=1}^3c_i= 0.
\end{align}

\begin{figure}[ht!]
	\leavevmode
	\centering
	\includegraphics[scale=1]{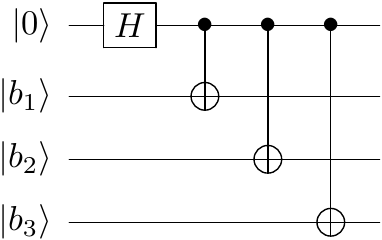}
	\caption{Quantum circuit for the operator $\hat{N}$.}
	\label{Ncircuit}
\end{figure}

Let us observe that the states in the pairs in Eq. \ref{Ipairs} are mutually negated. 
Furthermore, in each pair, the states have the same first binary digit. This suggests to 
consider an operator $\hat{N}$, which acts as follows:
\begin{equation}
 \hat{N}|0 b_1 b_2b_3 \rangle=\frac{\left(|0 b_1 b_2b_3  \rangle +|1 \bar{b}_1 \bar{b}_2\bar{b}_3  \rangle \right)}{\sqrt{2}},  
\end{equation}
generating from a given state $|0 b_1 b_2b_3 \rangle$ an equally weighted 
superposition of the state and its negation. The $b_1, b_2, b_3 \in \{ 0,1 \}$. 
Quantum circuit corresponding to the action of $\hat{N}$ can be constructed using 
combination of a single Hadamard gate and three CNOT gates. The quantum 
circuit is shown in Fig. \ref{Ncircuit}.

Employing the operator $\hat{N}$, the state $| \mathcal{I} \rangle$ can be expressed as,   
\begin{equation}
 | \mathcal{I} \rangle  = \hat{N}( |0\rangle  | \psi \rangle ) 
\end{equation}
where $|\psi \rangle $ is a 3-qubit state: 
\begin{equation}
	|\psi \rangle  = c_1 |011\rangle+c_2|101\rangle+c_3|110\rangle.   
\end{equation}

The task is now to find an operator $\hat{M}$, action o which is: 
\begin{equation}
	|\psi \rangle  = \hat{M}|000\rangle.
\end{equation}

\begin{figure}[ht!]
	\leavevmode
	\centering
	\includegraphics[scale=1]{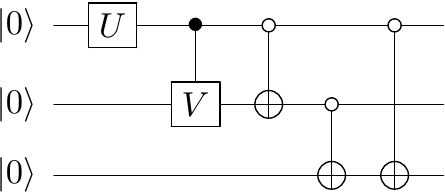}
	\caption{Quantum circuit for the operator $\hat{M}$.}
	\label{Mcircuit}
\end{figure}

One can find that the operator $\hat{M}$ is represented by the circuit presented in Fig. \ref{Mcircuit}.

In the circuit, the unitary operation $\hat{U}$, given by the special unitary matrix: 
\begin{equation}
U=\left(
\begin{array}{cc}
c_1 & \sqrt{|c_2|^2+|c_3|^2} \\
-\sqrt{|c_2|^2+|c_3|^2} & c_1^*
\end{array}	
\right),
\label{Umatrix}
\end{equation}
is performed first on the top qubit. Then, controlled-V  2-qubit gate is performed, 
where the special unitary matrix V is given by:  
\begin{equation}
V=\left(
\begin{array}{cc}
-\frac{c_2}{\sqrt{|c_2|^2+|c_3|^2}} & \frac{c_3^*}{\sqrt{|c_2|^2+|c_3|^2}} \\
-\frac{c_3}{\sqrt{|c_2|^2+|c_3|^2}} & -\frac{c_2^*}{\sqrt{|c_2|^2+|c_3|^2}}
\end{array}
\right).
\label{Vmatrix}	
\end{equation}
Finally, a sequence of three anti-CNOT gates, which allow to obtain deserved 
sequences of bits, are applied. 

\begin{figure}[ht!]
	\leavevmode
	\centering
	\includegraphics[scale=1]{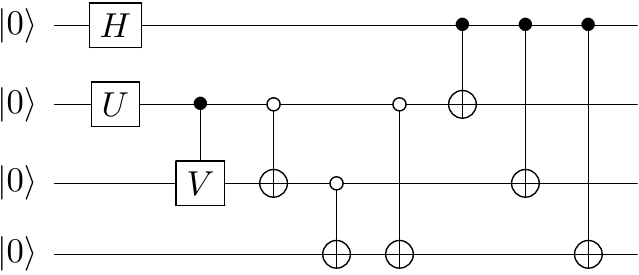}
	\caption{A quantum circuit corresponding to the operator $\hat{U}_{\mathcal{I}}$.}
	\label{QubitCircuit}
\end{figure}

Combining action of the operators $\hat{M}$ and $\hat{N}$ the general intertwiner 
state (\ref{IntertwinerState}) can now be written as: 
\begin{equation}
| \mathcal{I} \rangle  = \hat{U}_{\mathcal{I}} |0000\rangle = \hat{N}(\hat{\mathbb{I}} \otimes  \hat{M}) |0000\rangle.
\end{equation}
The corresponding quantum circuit is shown in Fig. \ref{QubitCircuit}. 

\section{Transpilation} \label{Sec:Transpilation}

Physical realizations of quantum computers impose restrictions on the types of 
quantum circuits which can be executed or implemented directly on a given 
quantum processor. In particular, the limitation is due to topology of couplings 
between the physical qubits. Because of this, \emph{transpilation} of the considered 
quantum circuit has to be performed, such that the circuit can be simulated on 
a given hardware.    

Here, we will consider transpilation of the quantum circuit shown in 
Fig. \ref{QubitCircuit} to the form compatible with the 5-qubit and 15-qubit 
quantum processors, made available by IBM via cloud computing platform \cite{IBM}. 

The transpilation concerns not only connectivity of the quantum processor 
but also the types of gates which are possible to execute. The Hadamard  
and CNOT gates are part of the standard IBM library. The anti-CNOT gate can 
be built utilizing the CNOT gate and two bit-flip $\hat{X}$ gates 
(corresponding to the Pauli matrix $X=\left(\begin{array}{cc} 0 & 1 \\ 
1 & 0 \end{array} \right)$): $(\hat{X} \otimes \hat{I}) \widehat{\text{CNOT}}(\hat{X} \otimes \hat{I})$.     
Furthermore, IBM utilizes the following gates: 
\begin{equation}
	U_1\left(\rho\right)=\left(
	\begin{array}{cc}
	1 & 0 \\
	0 & e^{i\rho}
	\end{array}
	\right),
\end{equation}
and 
\begin{equation}
	U_3\left(\theta,\phi,\lambda\right)=\left(
	\begin{array}{cc}
	\cos\left(\theta/2\right) & -e^{i\lambda}\sin\left(\theta/2\right) \\
	e^{i\phi}\sin\left(\theta/2\right) & e^{i\lambda+i\phi}\cos\left(\theta/2\right),
	\end{array}
	\right),
\end{equation}
which can be used to construct the operators $\hat{U}$ and $\hat{V}$. 
Namely, the operator $\hat{U}$ can be expressed as: 
\begin{equation}
	U=U_3\left(\theta_U,\phi_U,\lambda_U\right)X
\end{equation}
where the angles are
\begin{align}
	\theta_U&=2\arcsin\left(\sqrt{\frac{2}{3}}\sin\left(\theta/2\right)\right),\\
	\phi_U&=-\phi,\\
	\lambda_U&=\pi+\phi.
\end{align}
Similarly, the operator $\hat{V}$ can be written as:
\begin{equation}
	V=U_3\left(\theta_V,\phi_V,\lambda_V\right)XU_1\left(\rho_V\right),
\end{equation}
where 
\begin{align}
	\theta_V&= 2 \arcsin\left(\sqrt{ \frac{1}{2}+\frac{\sin\theta\cos\phi}{2\sqrt{3}\left(1-2/3\sin^2\theta\right)}}\right),\\
	\phi_V&=-\pi+\chi_+-\chi_-,\\
	\lambda_V&=\chi_-,\\
	\rho_V&=\pi-\chi_+.
\end{align}

\begin{figure}[ht!]
	\includegraphics[scale=0.4]{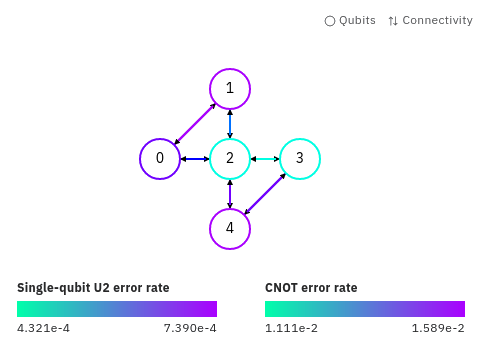}
	\caption{Connectivity of the IBM Q 5-qubit quantum processor (Yorktown). Obtained from IBM quantum cloud computing service \cite{IBM}.}
	\label{Yorktown}
\end{figure}

Let us now proceed to the topological considerations. In Fig. \ref{Yorktown}, 
connectivity of the 5-qubit IBM quantum processor is shown. 

\begin{figure}[ht!]
	\leavevmode
	\centering
	\includegraphics[scale=1]{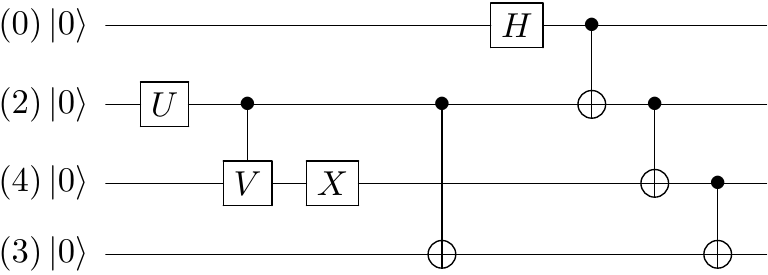}
	\caption{A quantum circuit corresponding to the operator $\hat{U}_{\mathcal{I}}$, compatible with the  
	 the 5-qubit IBM quantum processor Yorktown.}
	\label{TranspiledYorktown}
\end{figure}

The transpilated version of the circuit (\ref{QubitCircuit}) in agreement with the topology of the 
5-qubit quantum processor (Yorktown) is presented in Fig. \ref{TranspiledYorktown}.

\begin{figure}[ht!]
	\includegraphics[scale=0.4]{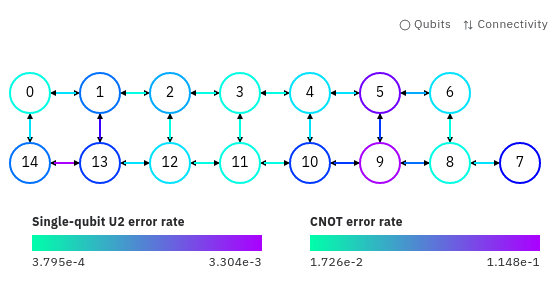}
	\caption{Connectivity of the IBM Q 15-qubit quantum processor (Melbourne). 
	Obtained from IBM quantum cloud computing service \cite{IBM}.}
	\label{Melbourne}
\end{figure}

In Fig. \ref{Melbourne} connectivity of the 15-qubit IBM quantum processor is shown. 

\begin{figure}[ht!]
	\leavevmode
	\centering
	\includegraphics[scale=1]{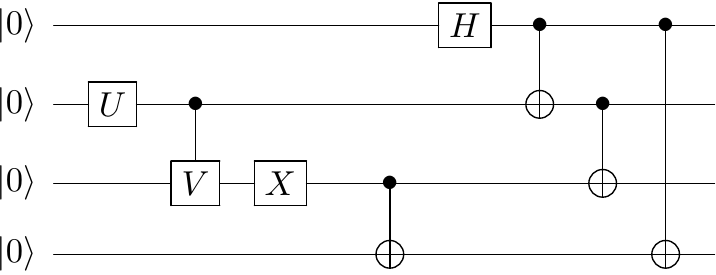}
	\caption{A quantum circuit corresponding to the operator $\hat{U}_{\mathcal{I}}$, compatible with the  
	 the 15-qubit IBM quantum processor Melbourne - version 1.}
	\label{TranspiledMelbourne_1}
\end{figure}

\begin{figure}[ht!]
	\leavevmode
	\centering
	\includegraphics[scale=0.9]{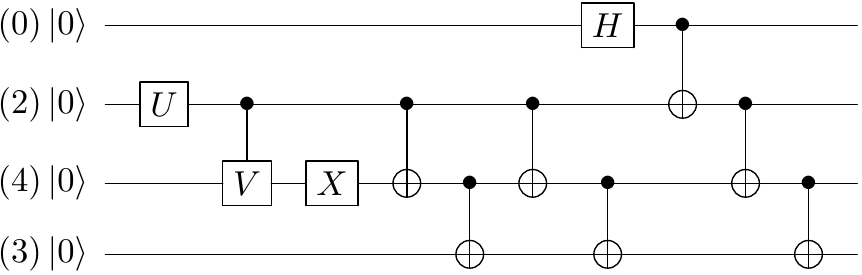}
	\caption{A quantum circuit corresponding to the operator $\hat{U}_{\mathcal{I}}$, compatible with the  
	 the 15-qubit IBM quantum processor Melbourne - version 2.}
	\label{TranspiledMelbourne_2}
\end{figure}

Two alternative versions of the transpilated circuit (\ref{QubitCircuit}), being in agreement with 
the topology of the 15-qubit quantum processor (Melbourne), are presented in 
Fig. \ref{TranspiledMelbourne_1} and Fig. \ref{TranspiledMelbourne_2}. 

\begin{figure}[ht!]
	\leavevmode
	\centering
	\includegraphics[scale=0.8]{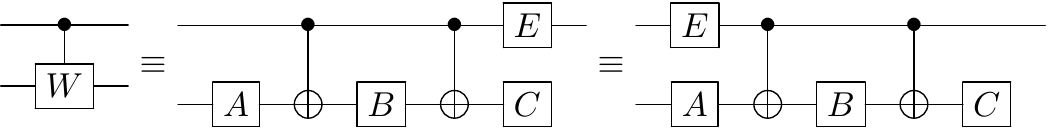}
	\caption{Control-W gate and its equivalent expressed with the use of single qubit gates and CNOT gates.}
	\label{ABC}
\end{figure}

One final issue is the controlled-V gate, which not necessary can be directly implemented. 
In that case, the 2-qubit gate can be expressed with the use of standard decomposition 
presented in Fig. \ref{ABC}, for a unitary operator $\hat{W}$ \cite{Barenco}. 

Here, $\hat{W}=\hat{V}\hat{S}$, where $\hat{V}$ is a special unitary operator and 
$\hat{S}=e^{i\delta} \hat{\mathbb{I}}$, with the phase $\delta \in \mathbb{R}$. The $E$ gate is given by the matrix 
$E=\left(\begin{array}{cc} 1&0\\ 0&e^{i\delta}\\ \end{array}\right)$. Furthermore, 
the involved gates $\hat{A}$, $\hat{B}$ and $\hat{C}$ are 1-qubit gates, satisfying conditions 
$\hat{C}\hat{B}\hat{A} = \hat{\mathbb{I}}$ and $\hat{C} \hat{X} \hat{B}\hat{X}\hat{A} = \hat{V}$.
In our case, because $V$ given by Eq. \ref{Vmatrix} is a special unitary matrix, we have 
$\delta=0$  so $\hat{W}=\hat{V}$ and matrix representations of the gates $A$, $B$ and $C$ are:
\begin{align}
	&A=\left(
	\begin{array}{cc}
	\cos \left(\frac{\rho}{4}\right) & \sin\left(\frac{\rho }{4}\right) \\
	e^{i (\chi_- -\chi_+)} \sin \left(\frac{\rho}{4}\right) & -e^{i (\chi_- -\chi_+ )}\cos\left(\frac{\rho }{4}\right) \\
	\end{array}
	\right), \nonumber   \\
	&B=\left(
	\begin{array}{cc}
	\cos \left(\frac{\rho }{4}\right) & e^{i\chi_+} \sin \left(\frac{\rho}{4}\right) \\
	\sin \left(\frac{\rho }{4}\right) & -e^{i\chi_+} \cos \left(\frac{\rho}{4}\right)\\	
	\end{array}
	\right), \nonumber   \\
	&C=\left(
	\begin{array}{cc}
	1 & 0 \\
	0 & e^{-i \chi_-} \\
	\end{array}
	\right),
\end{align}
where 
\begin{equation}
	\rho=2 \arccos\left(\sqrt{\frac{1}{2}\left(1-\frac{\sin\theta \cos \phi}{\sqrt{3}\left(1-\frac{2}{3}\sin^2\left(\frac{\theta}{2}\right)\right)}\right)}\right). 
\end{equation}
Furthermore, the gates can be constructed with use of the $U_3$ and $U_1$ gates as follows:
\begin{align}
	&A=U_3\left(\frac{\rho}{2},\chi_{-}-\chi_{+},\pi\right), \\
	&B=U_3\left(\frac{\rho}{2},0,\chi_{+}+\pi\right), \\
	&C=U_1\left(-\chi_{-}\right).
\end{align}

\section{Exemplary states} \label{Sec:Examples}

In this section we will simplify the obtained general quantum circuit shown 
in Fig. \ref{QubitCircuit} for the special cases of the intertwiner qubit basis 
states: $|0_s\rangle$ and  $|1_s\rangle$. This will allow to slightly reduce 
the general circuit, which is relevant from the perspective of quantum simulation, 
where the number of involved gates has to be minimized because of the 
issue of errors.  

\subsection{The state $|0_s\rangle$}

\begin{figure}[ht!]
	\leavevmode
	\centering
	\includegraphics[scale=1]{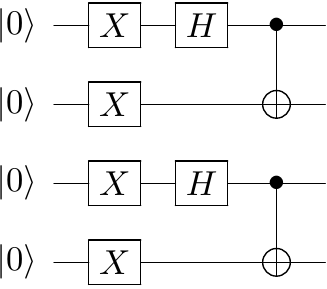}
	\caption{Quantum circuit for the $|0_s\rangle$ state discussed in Ref. \cite{Mielczarek:2018jsh}.}
	\label{0sStateOld}
\end{figure}

The quantum circuit for the $|0_s\rangle$ state has already been a subject of 
investigation in Ref. \cite{Mielczarek:2018jsh} and is shown in Fig. \ref{0sStateOld}.

Here, we will present an alternative construction of the state, starting from the general circuit 
shown in Fig. \ref{QubitCircuit}. Taking $\theta=0$, we find that the coefficients: 
\begin{equation}
	c_1 =0, \ \
	c_2 =\frac{1}{\sqrt{2}}, \ \
	c_3 =-\frac{1}{\sqrt{2}}.
\end{equation}

In consequence, the $\hat{U}$ and $\hat{V}$ operators (see Eqs. \ref{Umatrix} and \ref{Vmatrix}) 
are now represented by the following matrices:   
\begin{equation}
U=\left(
\begin{array}{cc}
0 & 1 \\
-1 & 0
\end{array}	
\right)
\end{equation}
and
\begin{equation}
V=\frac{1}{\sqrt{2}}\left(
\begin{array}{cc}
-1 & -1 \\
1 & -1
\end{array}	
\right).
\end{equation}

\begin{figure}[ht!]
	\leavevmode
	\centering
	\includegraphics[scale=1]{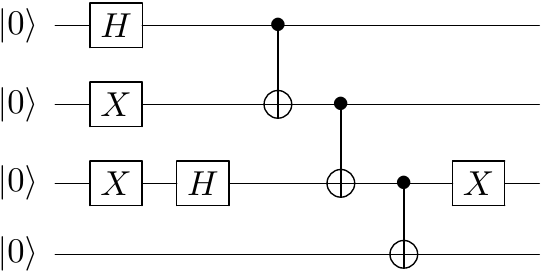}
	\caption{Quantum circuit for the $|0_s\rangle$ state.}
	\label{0sStateNew}
\end{figure}

This allows to reduce the circuit from Fig. \ref{QubitCircuit} to the one presented
in Fig. \ref{0sStateNew}.

\subsection{The state $|1_s\rangle$}

For the state $|1_s\rangle$, we take $\theta=\pi$, which reduces the coefficients (\ref{c1}), (\ref{c2}) and (\ref{c3}) to:
\begin{equation} 
c_1=\sqrt{\frac{2}{3}}, \ \ 
c_2=-\frac{1}{\sqrt{6}}, \ \
c_3=-\frac{1}{\sqrt{6}},
\end{equation}
such that the $U$ and $V$ matrices are 
\begin{equation}
U=\sqrt{\frac{2}{3}}\left(
\begin{array}{cc}
1 & \frac{1}{\sqrt{2}} \\
-\frac{1}{\sqrt{2}} & 1
\end{array}	
\right),
\end{equation}
and
\begin{equation}
V=\frac{1}{\sqrt{2}}\left(
\begin{array}{cc}
1 & -1 \\
1 & 1
\end{array}	
\right).
\end{equation}

\begin{figure}[ht!]
	\leavevmode
	\centering
	\includegraphics[scale=1]{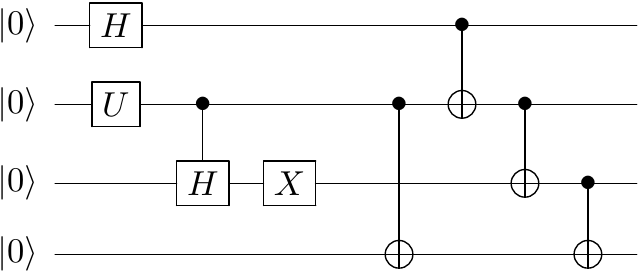}
	\caption{Quantum circuit for the $|1_s\rangle$ state.}
	\label{1sState}
\end{figure}

This allows to reduce the circuit from Fig. \ref{QubitCircuit} to the one presented
in Fig. \ref{1sState}.

\section{Quantum simulations} \label{Sec:Simulations}

The quantum circuits for a single intertwiner qubit introduced in the previous 
sections represent unitary operators acting in 16 dimensional Hilbert space, being 
a tensor product of four $\mathcal{H}_{1/2}$ Hilbert spaces. Such a case is 
easy to handle wit the use of classical computer. However, the difficulty came 
when more complex systems are considered. In our case, four qubits are 
needed to define a single intertwiner qubit. Therefore, a spin network with 
$N$ four-valent nodes requires $4N$ logical qubits. The corresponding Hilbert 
space has dimension $\text{dim} \otimes_{i=1}^{4N} \mathcal{H}_i  = 2^{4N}$. 
In case of a general quantum circuit, classical simulations of the systems 
with $N\sim 20$  ($\sim 80$ logical qubits) is already beyond the reach of 
any currently existing classical supercomputer \cite{Chen}. On the other hand, 
(noisy) quantum computers with the number of qubits $\sim 50$ already exist 
and the ones with $\sim 100$ are under development (see e.g. \cite{IBM,Google,Rigetti,Ionq}). 
This prognosis that simulations of spin networks with $N\sim 20$ and more will 
become feasible in the coming years (see also discussion in Ref. \cite{Mielczarek:2018jsh}). 
However, as we will already see while considering 15 qubit quantum chip, the 
issue of errors reduction remains to be a challange even in processors with 
over a dozen of qubits. Furthermore, we have to emphasize that the superconducting 
quantum computers are characterized by relatively short coherence times, 
which limits depth of the quantum circuits which can be simulated successfully.    

Here, we will present results of simulations of exemplary states of the intertwiner 
qubit performed on 5-qubit (Yorktown) and 15-qubit (Melbourne) IBM superconducting 
quantum processors, topologies of which are shown in Fig. \ref{Yorktown} and Fig. 
\ref{Melbourne} respectively. In the figures, errors of the particular qubits at the
time of simulations are also presented.  

The six representative states which are considered are: $|0_s\rangle$, $|1_s\rangle$, 
$|+\rangle:=\frac{|0_s\rangle+|1_s\rangle}{\sqrt{2}}$, $|-\rangle:=\frac{|0_s\rangle-|1_s\rangle}{\sqrt{2}}$, 
$|\leftturn\rangle:=\frac{|0_s\rangle-i|1_s\rangle}{\sqrt{2}}$ and $|\rightturn\rangle:=\frac{|0_s\rangle+i|1_s\rangle}{\sqrt{2}}$.
The $|0_s\rangle$ and  $|1_s\rangle$ states correspond to the points on the north and south 
pole of the Bloch sphere correspondingly. The remaining four states are the points located 
at the equator of Bloch sphere, and are evenly distributed with the polar angle difference 
$\Delta \phi = \frac{\pi}{2}$. The considered states have direct physical interpretation if they 
are referred to light. Namely, if $|0_s\rangle$, $|1_s\rangle$ are horizontal $(|\text{H}\rangle)$ 
and vertical $(|\text{V}\rangle)$ linear polarization states of a photon respectively, then 
the $|+\rangle$ and  $|-\rangle$ are $\pm \frac{\pi}{4}$ linear polarization states.  
The $|\leftturn\rangle$ is a left-hand circular polarization state and $|\rightturn\rangle$ is a 
right-hand circular polarization state, which justifies the applied notation. Furthermore, 
the $|\leftturn\rangle$ and $|\rightturn\rangle$ are also eigenstates of the volume operator. 
Namely, based on (\ref{Vplus}) and (\ref{Vminus}) one can see that:
\begin{equation}
|\leftturn\rangle = |V_+\rangle \ \text{and} \ \ |\rightturn\rangle = |V_-\rangle.
\end{equation} 

In the simulations, a sequence of 10 computational rounds each containing 1024 shots 
was performed for every of the investigated states. The simulations were performed on 
both the 5-qubit Yorktown quantum processor and 15-qubit Melbourne quantum processor. 
Topologies of the processors together with the errors (single-qubit and CNOT 2-qubit gate) 
at the time of simulations are depicted in Fig. \ref{Yorktown} and Fig. \ref{Melbourne}. 
The obtained averaged measured probabilities of the basis states for each of the states are 
shown in Fig. \ref{ProbabilitiesStates}. Detailed numerical results of the simulations can 
be found in Appendix A.
\begin{figure*}
	\includegraphics[scale=0.8]{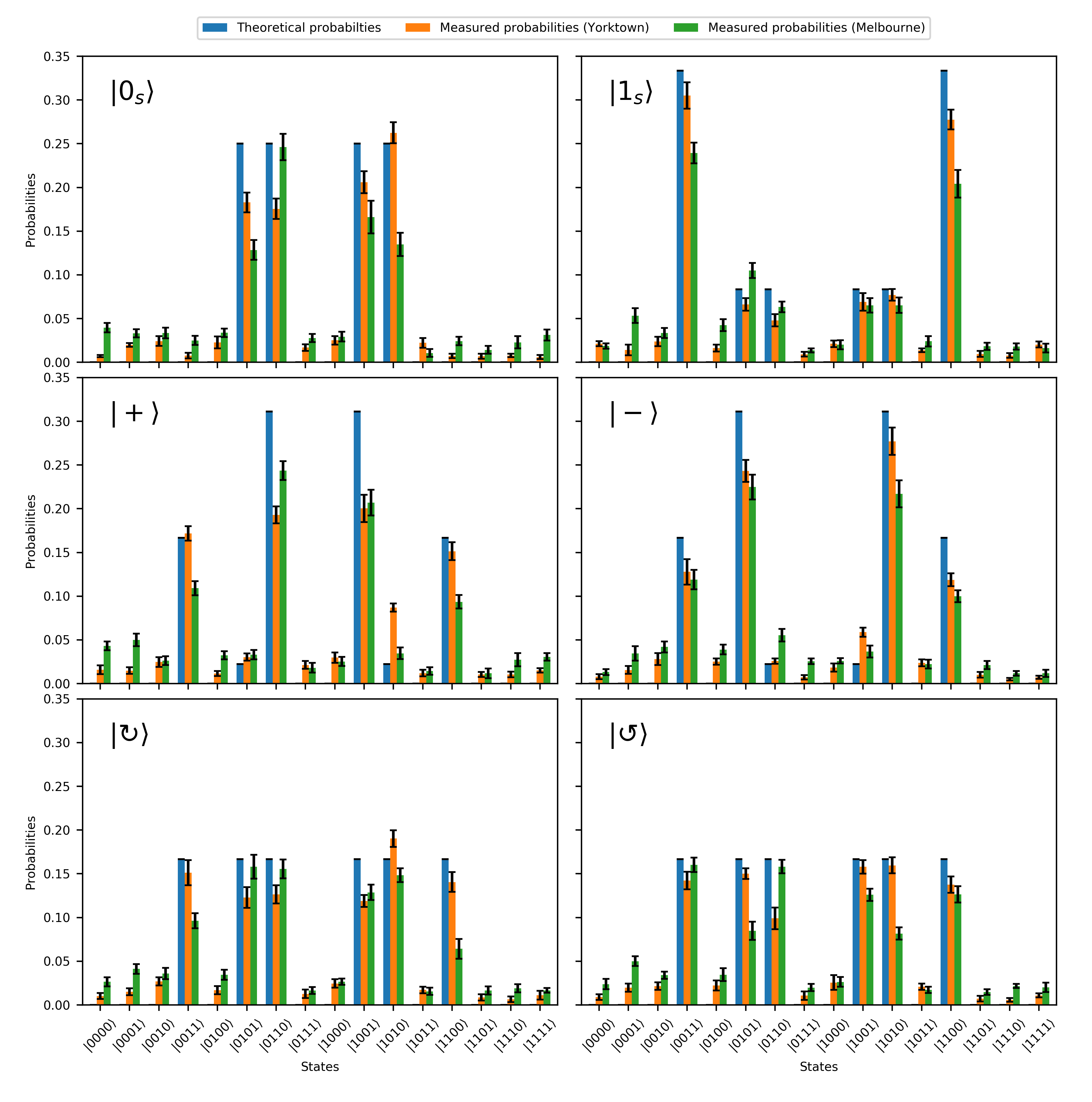}
	\caption{Measured and theoretical probabilities for the six representative states.}
	\label{ProbabilitiesStates}
\end{figure*} 
 
\begin{figure}
	\includegraphics[scale=0.35]{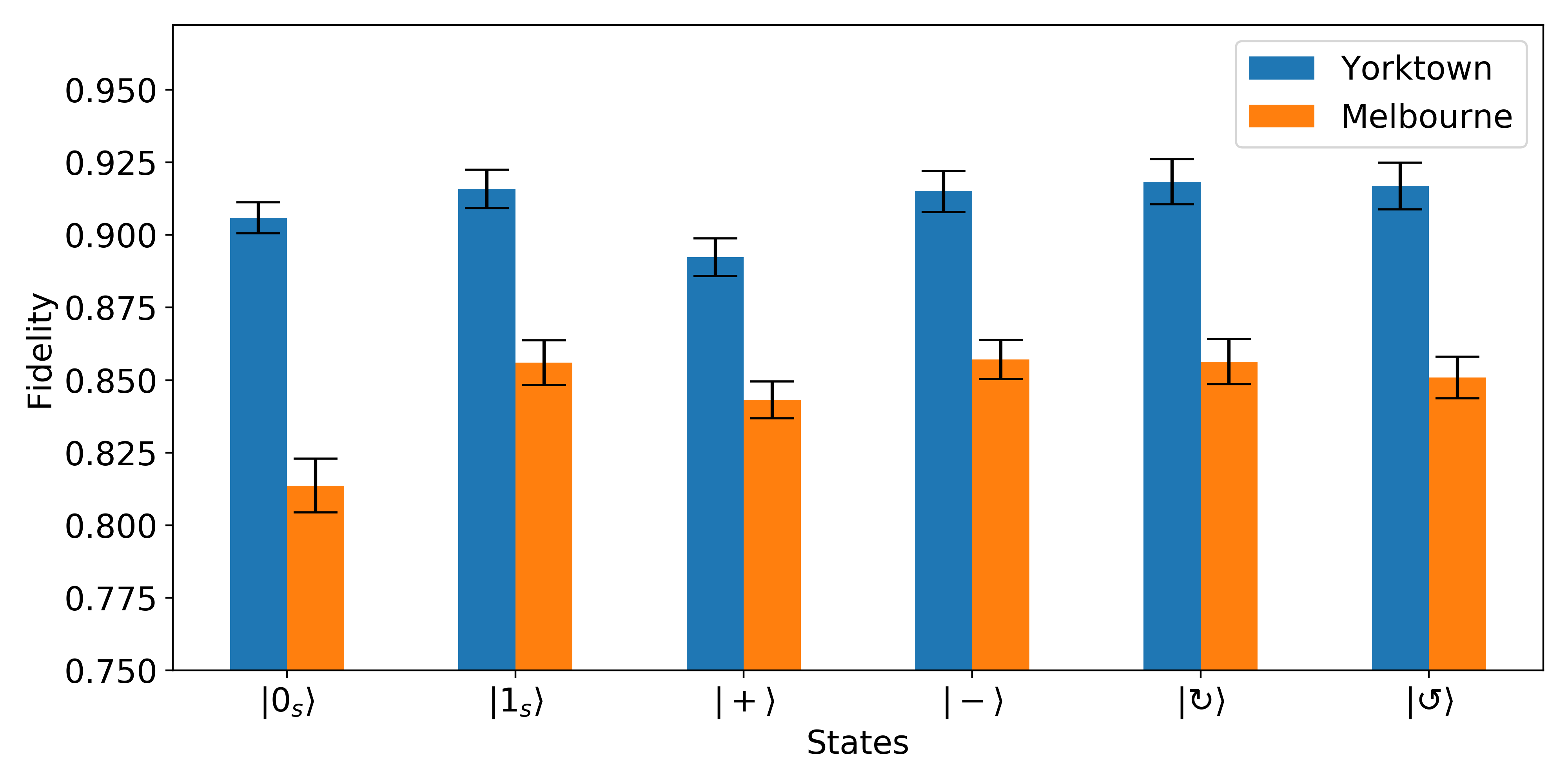}
	\caption{Fidelity for the six representative states of the intertwiner qubit 
	generated on 5-qubit (Yorktown) and 15-qubit (Melbourne) IBM quantum 
	computers.}
	\label{FidelitiesStates}
\end{figure}

In order to quantify difference between the measured states and theoretical values 
we use the classical fidelity function (Bhattacharyya distance): 
 \begin{equation}
	F\left(p,q\right)=\sum_i\sqrt{p_iq_i}.
\end{equation}
More detailed analysis would require quantum tomography of the states. However, 
consideration of the classical fidelity function is sufficient for our purpose. The obtained 
fidelities are collected in Table \ref{FidelityTable}, and presented in Fig. \ref{FidelitiesStates}.
\begin{table}
	\[
	\begin{array}{|c|c|c|}
	\hline
	\text{State}&\text{Yorktown}&\text{Melbourne}\\
	\hline
	|0_s\rangle & 0.906\pm0.005 & 0.814\pm0.009\\
	|1_s\rangle & 0.916\pm0.007 & 0.856\pm0.008\\
	|+\rangle & 0.892\pm0.007 & 0.843\pm0.006\\
	|-\rangle & 0.915\pm0.007 & 0.857\pm0.007\\
	|\circlearrowright\rangle & 0.918\pm0.008 & 0.856\pm0.008\\
	|\circlearrowleft\rangle & 0.917\pm0.008 & 0.851\pm0.007\\
	\hline
	\end{array}
	\]
	\caption{Values of fidelity for the six representative states under consideration.}
	\label{FidelityTable}
\end{table}

In case of the 5-qubit chip, the fidelities of the obtained states reach the level of 
$F\approx 90$ \%. This is a significant increase comparing to the fidelity $F\approx 71$ \%
of the state $|0_s\rangle$ obtained in Ref. \cite{Mielczarek:2018jsh}. Furthermore, 
simulations of the same states performed on the 15-qubit chip are at the level 
$F\approx 85$ \%. There is no significant difference in the fidelities depending 
on which state is considered. 

For comparison, the fidelities obtained in Ref. \cite{Li:2017gvt} 
(employing molecular quantum computer) are better than those obtained here. 
However, in our approach the superconducting chip was used, which despite 
of being more noisy, gives better perspective for scaling to more complicated
cases. On the other hand, in Ref. \cite{Zhang:2020lwi} spin foam vertex amplitude 
(composed of five intertwiner qubits) has been simulated with fidelity $0.832\pm0.005$, 
which is lower than almost all of the fidelities obtained here. This state had, 
however, more complex circuit structure than the circuits considered here. 
Furthermore, the results were possible to obtain because the simulations 
were performed directly on the intertwiner qubits.  
	
The states' imperfectness is mostly because of the errors associated 
with three factors: preparation of the initial state, implementation of the 
quantum gates, and readout. The errors (mainly corresponding to the two-qubit 
gates) have been significantly reduced over recent years. The errors of 
gates are shown in Fig.\ref{Yorktown} and in Fig.\ref{Melbourne}. 

In particular, in case of the Yorktown processor, the error of the single-qubit 
$U_2\left(\lambda,\phi\right)=\frac{1}{\sqrt{2}}\left(\begin{array}{cc}
1&-e^{i\lambda}\\
e^{i\phi}&e^{i\left(\phi+\lambda\right)}
\end{array}\right)$ gate is in range between $4.3\cdot 10^{-4}$ and $7.4\cdot 10^{-4}$. 
For the CNOT gate the error rate is between $1.1\cdot 10^{-2}$ and $1.6\cdot 10^{-2}$. 
For the Melbourne processor these errors are  between $3.8\cdot 10^{-4}$ and $3.3\cdot 10^{-3}$ 
and between $1.7\cdot 10^{-2}$ and $1.1\cdot 10^{-1}$ respectively. The single-qubit $U_2$ 
error rates and CNOT error rates have been measured using randomized benchmarking 
procedure \cite{benchmarking}.  Despite of the considerable hardware improvement, quantum 
error corrections codes \cite{Lidar} can also be implemented to further reduce the errors. 
However, this can be achieved only in a quantum chip with a sufficiently high number 
of physical qubits. This does not concern the currently available solutions. However, a 
set of methods called \emph{error mitigation} \cite{mitigation} can also be used to further 
improve the results. We plan to apply these methods in our future research.

\section{Transition amplitudes} \label{Sec:Amplitudes}

The results presented so far can be applied to evaluate transition amplitudes 
between states of spin networks (of fixed topology), representing different 
quantum geometries. In case of quantum gravity, and other quantum constrained 
systems, the subtlety is that the states under consideration have to be appropriately 
projected onto the physical Hilbert space $\mathcal{H}_{\text{phys}}$. In consequence, 
while some kinematical states $|\psi_1\rangle, |\psi_2\rangle \in \mathcal{H}_{\text{kin}}$ 
are considered, the corresponding transition amplitude has the following form:  
\begin{equation}
	A\left(\psi_1,\psi_2\right) :=\langle \psi_2|\hat{P}|\psi_1\rangle,
	\label{Amplitude}
\end{equation}
where $\hat{P}$ is a non-unitary, but Hermitian ($\hat{P}^{\dagger}=\hat{P}$) and idempotent 
($\hat{P}^2=\hat{P}$), projector operator.  In consequence, the $\hat{P}$ cannot be associated 
with a unitary quantum circuit.  On the other hand, in the context of quantum 
computing, action of the projection operators is associated with quantum measurements.

In case when more than one constraint is involved, as in the case of gravity, the projection 
operator is a composition of projection operators for the individual constraints:
\begin{equation}
\hat{P} = \hat{P}_1 \circ \hat{P}_2 \circ \dots  \circ \hat{P}_m,
\end{equation} 
where $m$ is the number of constraints. 

In LQG, the constraint are grouped into the three types: Gauss constraint, Diffeomeorphism constraint 
(vector constraint) and the Hamiltonian constraint (scalar constraint). Here, we will focus our attention 
on the case of the Gauss constraint, which is employed in the construction of the spin network spates. 
The vector constraint is on the other hand satisfied just by the graph structure of the spin network, so 
it is satisfied by construction. The scalar constraint is the most difficult to satisfy and we are not going
to discuss it here.  However, quantum computing methods provide some new possibilities to address 
the problem \cite{Mielczarek:2018ttq}. 

In order to compute the amplitude (\ref{Amplitude}) with the use of quantum circuits, let 
us consider operators $\hat{U}_{\psi_1}$ and $\hat{U}_{\psi_2}$, defined such that 
$|\psi_1\rangle=\hat{U}_{\psi_1}|{\bf 0} \rangle$ and $|\psi_2\rangle=\hat{U}_{\psi_2}|{\bf 0} \rangle$.
The $|{\bf 0} \rangle$ is an initial state of the quantum register, which in case of the 
spin network with $N$ four-valent nodes is $|{\bf 0} \rangle = \otimes_{i=1}^{4N}|0\rangle$. 
In consequence, the transition amplitude (\ref{Amplitude}) takes the form: 
\begin{equation}
	\langle \psi_2|\hat{P}|\psi_1\rangle = \langle {\bf 0} |\hat{U}^{\dagger}_{\psi_2} 
	\hat{P}\hat{U}_{\psi_1}|{\bf 0}\rangle.
	\label{Amplitude2}
\end{equation}

Because $\hat{P}$ is a non-unitary operator, the operator $\hat{U}^{\dagger}_{\psi_2} 
\hat{P}\hat{U}_{\psi_1}$ cannot be represented by a standard quantum circuit. There 
is, however, a special case when at least one of the states  $|\psi_1\rangle$ and $|\psi_2\rangle$ 
is invariant under the action of the projection operator $\hat{P}$. Then, for the Gauss constraint, 
this means that at least one of the states is a superposition of spin network states. 

\begin{figure}[ht!]
	\leavevmode
	\centering
	\includegraphics[scale=1]{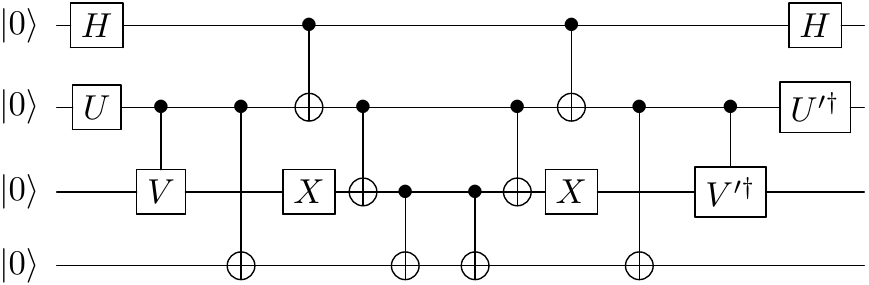}
	\caption{Quantum circuit for the transition amplitude operator $\hat{U}=\hat{U}^{\dagger}_{\mathcal{I}'}\hat{U}_{\mathcal{I}}$ between two arbitrary intertwiner states $|\mathcal{I}\rangle$ and  $|\mathcal{I}'\rangle$.}
	\label{fig:amplitude_before}
\end{figure}

Let us examine such possibility first for the case of a single node of a spin network. 
In that case, for the intertwiner qubit, the projection operator associated with the Gauss 
constraint takes the form:
\begin{equation}
\hat{P}_{\text{G}} = |0_s \rangle \langle 0_s | + |1_s \rangle \langle 1_s |. 
\label{GaussProjection1}
\end{equation}   
Then, if e.g. $|\psi_1\rangle$ is a state of intertwiner qubit, it can be expressed 
as follows:
 \begin{equation}
| \psi_1 \rangle =\hat{U}_{\psi_1}|{\bf 0} \rangle= \cos(\theta_1/2) |0_s\rangle
+e^{i\phi_1} \sin(\theta_1/2)|1_s\rangle,
\end{equation} 
where now $|{\bf 0} \rangle = |0000\rangle$.  It is straightforward to show that 
$\hat{P}_{\text{G}}| \psi_1 \rangle = | \psi_1 \rangle$ and, in consequence, in the 
considered case, the transition amplitude (\ref{Amplitude2}) reduces to 
\begin{equation}
	\langle \psi_2|\hat{P}_{\text{G}}|\psi_1\rangle = 
	\langle {\bf 0} |\hat{U}^{\dagger}_{\psi_2}\hat{U}_{\psi_1}|{\bf 0}\rangle.
	\label{Amplitude3}
\end{equation}
Therefore, unitary operator $\hat{U}:=\hat{U}^{\dagger}_{\psi_2}\hat{U}_{\psi_1}$ can be introduced,
which can be associated with a quantum circuit. For transition between two intertwiner qubit states, 
the $| \psi_1 \rangle = |\mathcal{I}\rangle$ and $| \psi_2 \rangle = |\mathcal{I}'\rangle$, the quantum 
circuit corresponding to the operator $\hat{U}=\hat{U}^{\dagger}_{\psi_2}\hat{U}_{\psi_1}$ is shown in 
Fig. \ref{fig:amplitude_before}.

The $U$ and $V$ are matrices associated with the state $|\mathcal{I}\rangle$ 
and $U'$ and $V'$ are associated with $|\mathcal{I'}\rangle$, in accordance to 
the circuit presented in Fig. \ref{QubitCircuit}.

\begin{figure}[ht!]
	\leavevmode
	\centering
	\includegraphics[scale=1]{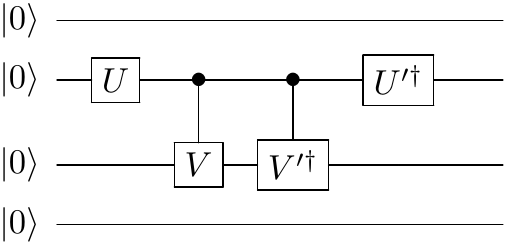}
	\caption{Simplified quantum circuit for the transition amplitude operator 
	$\hat{U}=\hat{U}^{\dagger}_{\mathcal{I}'}\hat{U}_{\mathcal{I}}$ between two 
	arbitrary intertwiner states $|\mathcal{I}\rangle$ and  $|\mathcal{I}'\rangle$.}
	\label{fig:amplitude_two_I}
\end{figure}

Using the fact that $\hat{X}$, $\hat{H}$ and $\widehat{\text{CNOT}}$ are unitary 
operators, the circuit shown in Fig. \ref{fig:amplitude_before} can be reduced to 
form presented in Fig. \ref{fig:amplitude_two_I}.

Therefore, only two qubits contribute non-trivially to the transition amplitudes 
$\langle \mathcal{I}'|\mathcal{I}\rangle$.

The above discussion can be extended to general superpositions of 4-valent spin 
network constructed with $N$ intertwiner qubits. Such a state can be written as 
\begin{equation}
|\psi \rangle =  \sum_{k_{1,s}\dots k_{N,s}  \in \{0,1\}}   c_{k_{1,s}, \dots, k_{N,s}}   \otimes_{i}^N  |k_{i,s}\rangle,
\label{GeneralSpinNetworkState}
\end{equation}
where $ |k_{i,s}\rangle$ is basis state of a $i-$the intertwiner qubit. The generalized version 
of Eq. \ref{GaussProjection1} to the case of $N$ intertwiner qubits is: 
\begin{align}
\hat{P}_{\text{G}} &= \otimes_{i=1}^N( |0_{i,s} \rangle \langle 0_{i,s}| + |1_{i,s} \rangle \langle 1_{i,s} |) \nonumber \\ 
                   &= \sum_{k_{1,s}\dots k_{N,s}  \in \{0,1\}}  |k_{1,s}\dots k_{N,s} \rangle \langle k_{1,s}\dots k_{N,s}|.
\label{GaussProjection2}
\end{align}
Direct action of the operator (\ref{GaussProjection2}) onto (\ref{GeneralSpinNetworkState})
confirms that $\hat{P}_{\text{G}}  |\psi \rangle =|\psi \rangle$. Therefore, always if at least one of 
the states in the transition amplitude $\langle \psi_2|\hat{P}_{\text{G}}|\psi_1\rangle$ is of the form 
of Eq. \ref{GeneralSpinNetworkState}, the transition amplitude reduces to 
$\langle \psi_2|\psi_1\rangle$ and quantum circuit corresponding to 
$\hat{U}=\hat{U}^{\dagger}_{\psi_2}\hat{U}_{\psi_1}$ can be introduced. As already discussed 
in Ref. \cite{Mielczarek:2018jsh}, action of this operator on the initial state of quantum 
register of $N$ intertwiner qubits can be written as 
$\hat{U}|{\bf 0}\rangle=\sum_{i=0}^{2^{4N}-1} a_i|{\bf i}\rangle$, where $|{\bf i}\rangle$ is a 
basis state in the $2^{4N}$ dimensional Hilbert space of the system. With the use of this, 
the transition amplitude (\ref{Amplitude}) can be written as: 
\begin{equation}
\langle \psi_2|\psi_1\rangle =\langle {\bf 0} |\hat{U}^{\dagger}_{\psi_2} \hat{U}_{\psi_1}|{\bf 0}\rangle=a_0,
\end{equation}
where $a_0 \in \mathbb{C}$ is the amplitude of $|{\bf 0}\rangle$ state in the final state 
obtained by evaluation of the quantum circuit. In practice, the probability $P_0=|a_0|^2$ is 
determined, unless tomography of the final quantum state if performed.  

\section{Maximally Entangled Spin Networks}  \label{Sec:EntangledNetwork}

The spin networks are built from holonomies, which from the quantum mechanical 
viewpoint, are unitary maps between two Hilbert spaces, associated with the endpoints
of a given curve $\lambda \in[0,1]  \rightarrow  e(\lambda) \in \Sigma$, where $\lambda$
is an affine parameter which parametrises the curve. Let us denote endpoint as $s=e(0)$ 
(source) and  $t=e(1)$ (target). Then, we can introduce the source and target Hilbert spaces 
$\mathcal{H}_s$ and $\mathcal{H}_t$ between which the holonomy is mapping. 

The relation between quantum entanglement and the spin networks was a 
subject of investigation for over a decade \cite{Donnelly:2016auv,Donnelly:2008vx,Feller:2017jqx}. 
This, especially, concerned understanding of the Bekenstein-Hawking formula in terms
of von Neumann entanglement entropy. However, it became evident only recently that a single 
$SU(2)$ holonomy is associated with a maximally entangled state \cite{Livine:2017fgq,Czech:2019vih,Mielczarek:2019srn}: 
\begin{align}
| \Psi \rangle := \frac{1}{\sqrt{2j+1}} h^*_{IJ} | I \rangle_s | J \rangle_t \in \mathcal{H}_s \otimes \mathcal{H}_t\,,
\label{PsiQubit}
\end{align}
where $h_{IJ}$ are matrix elements of the ${\rm SU}(2)$ holonomy.  The indices $I,J =0,1, \dots, 2j$, 
where $j$ labels irreducible representation of the SU(2) group. In the case of fundamental ($j=1/2$) 
representation, the (\ref{PsiQubit}) reduces to  
\begin{align}
| \mathcal{E}_l\rangle := \frac{1}{\sqrt{2}} h^*_{IJ} | I \rangle_{s,l} | J \rangle_{t,l}\,,
\label{MaxEnt}
\end{align} 
for a given link $l$ of a spin network, and $I,J =0,1$. The state is an example of maximally 
entangled state, in the sense of maximization of the mutual quantum information. 
Then, the total state for a graph can be written as: 
\begin{equation}
| \mathcal{E}\rangle = \bigotimes_l | \mathcal{E}_l\rangle, 
\end{equation}
where the tensor products runs over all links of the graph. The state introduced in this way, 
in general, does not satisfy the Gauss constraint. Therefore, in order express the state as a 
superposition of spin networks states, an appropriate projection has to be applied.  We define 
such state as \emph{maximally entangled spin network} (MESN) state:
\begin{equation}
| \text{MESN} \rangle := \hat{P}_{\text{G}} \bigotimes_l | \mathcal{E}_l \rangle.   
\end{equation}
It has to be emphasized that while the state is built out of maximally entangled pairs, 
the $\hat{P}_{\text{G}}$ projection is affecting the entanglement properties of the resulting 
state. However, in a deserved way. Namely, the construction of the MESN state is 
analogous to the way in which Projected Entangled Pair States (PEPS) \cite{PEPS1,PEPS2} 
tensor networks \cite{Orus:2013kga,Biamonte} are introduced. The projection onto a 
singlet state performed in the case PEPS tensor networks is just imposing the Gauss 
constraint in the case of spin networks. One of the important properties of the PEPS 
tensor networks is that they satisfy area-law scaling of the entanglement entropy \cite{Orus:2013kga}.
This is relevant from the viewpoint of utilizing MESN states in description of gravitational 
systems. In particular, this concerns black holes for which the Bekenstein-Hawking 
area law $S_{\text{BH}} = \frac{A}{4l^2_{\text{Pl}}}$, is satisfied. Furthermore, because 
of the holographic nature of the Gravity/Entanglement duality, studies of the MESN states 
may contribute to our better understanding of the conjecture. 

An example of the maximally entangled state (\ref{MaxEnt}) is the 2-qubit singlet state  
\begin{equation}
| \mathcal{E}_l \rangle=\frac{1}{\sqrt{2}}\left(|01\rangle-|10\rangle\right),
\label{BellState}
\end{equation}
which, based on Eq. \ref{MaxEnt}, corresponds to the following holonomy:
\begin{align}
h = \left( \begin{array}{cc} 0  & 1 \\ -1 & 0 \end{array} \right) =  i \sigma_y 
= e^{ i \frac{\pi}{2} \sigma_y}.
\label{HoloExample}
\end{align}
The state has been used to construct states of spin networks in Refs.  
\cite{Li:2017gvt,Mielczarek:2018jsh} and we will examine more properties 
of such a choice in the next two sections. 

Despite of certain similarities, the state introduced in this section differs from 
the \emph{Bell-network states} recently studied in Refs. \cite{Baytas:2018wjd,Bianchi:2018fmq}.
In that case, the Bell states (\ref{BellState}) and other maximally entangled states have 
been utilized, however, in that case Schwinger representation of the SU(2) group is used, 
such that at both source and target two copies of the bosonic Hilbert space 
are defined.  In such case, the Bell state for a given link is introduced by action of a 
squeezing operator on the four harmonic oscillators, which is different from the 
approach presented here.  

Below, we consider two examples of spin networks: monopole and 
dipole spin networks. Despite their simplicity, the elementary spin networks 
may have physical relevance. Namely, they can be considered as a cosmological 
approximation of spatial geometry. In particular, the dipole spin network represents 
minimal triangulation of a 3-sphere, i.e., two tetrahedra glued along each face. 
This configuration describes a non-homogeneous quantum universe. This 
observation has been broadly explored in the context of \emph{spin foam cosmology} 
\cite{Rovelli:2008aa, Borja:2010gn, Bianchi:2010zs}. Moreover, the quantum 
tetrahedra considered here find application in the framework of Group Field 
Theory \cite{Oriti:2016qtz}.

\section{Monopole Spin Network} \label{Sec:Monopole}

The simplest non-trivial example of a spin network is the case of a monopole with 
a single node. In order to construct the maximally entangled spin network state for 
such a case, let us rewrite Eq. \ref{BellState} for a link connecting $i$-th and $j$-th 
qubits as: 
\begin{equation}
	|\mathcal{E}_{ij}\rangle=\frac{1}{\sqrt{2}}\left(|0_i1_j\rangle-|1_i0_j\rangle\right).
\end{equation}

At the single node of the monopole graph, four links meet and in consequence there 
are three different possibilities to pair the qubits by two holonomies. The cases 
correspond to the following states: 
\begin{align}
|\mathcal{E}_{0123}\rangle &:=|\mathcal{E}_{01}\rangle|\mathcal{E}_{23}\rangle=|0_s\rangle \nonumber \\
& =  |\mathcal{I}\left(\theta=0,\phi \right)\rangle , \\
|\mathcal{E}_{0213}\rangle &:=|\mathcal{E}_{02}\rangle|\mathcal{E}_{13}\rangle=\frac{1}{2}|0_s\rangle+\frac{\sqrt{3}}{2}|1_s\rangle \nonumber \\
& = |\mathcal{I}\left(\theta={2\pi}/{3},\phi=0\right)\rangle,   \\
|\mathcal{E}_{0312}\rangle &:=|\mathcal{E}_{03}\rangle|\mathcal{E}_{12}\rangle=-\frac{1}{2}|0_s\rangle+\frac{\sqrt{3}}{2}|1_s\rangle \nonumber \\
& = -|\mathcal{I}\left(\theta={2\pi}/{3},\phi=\pi\right)\rangle.
\end{align}

\begin{figure}[ht!]
	\centering
	\includegraphics[scale=0.25]{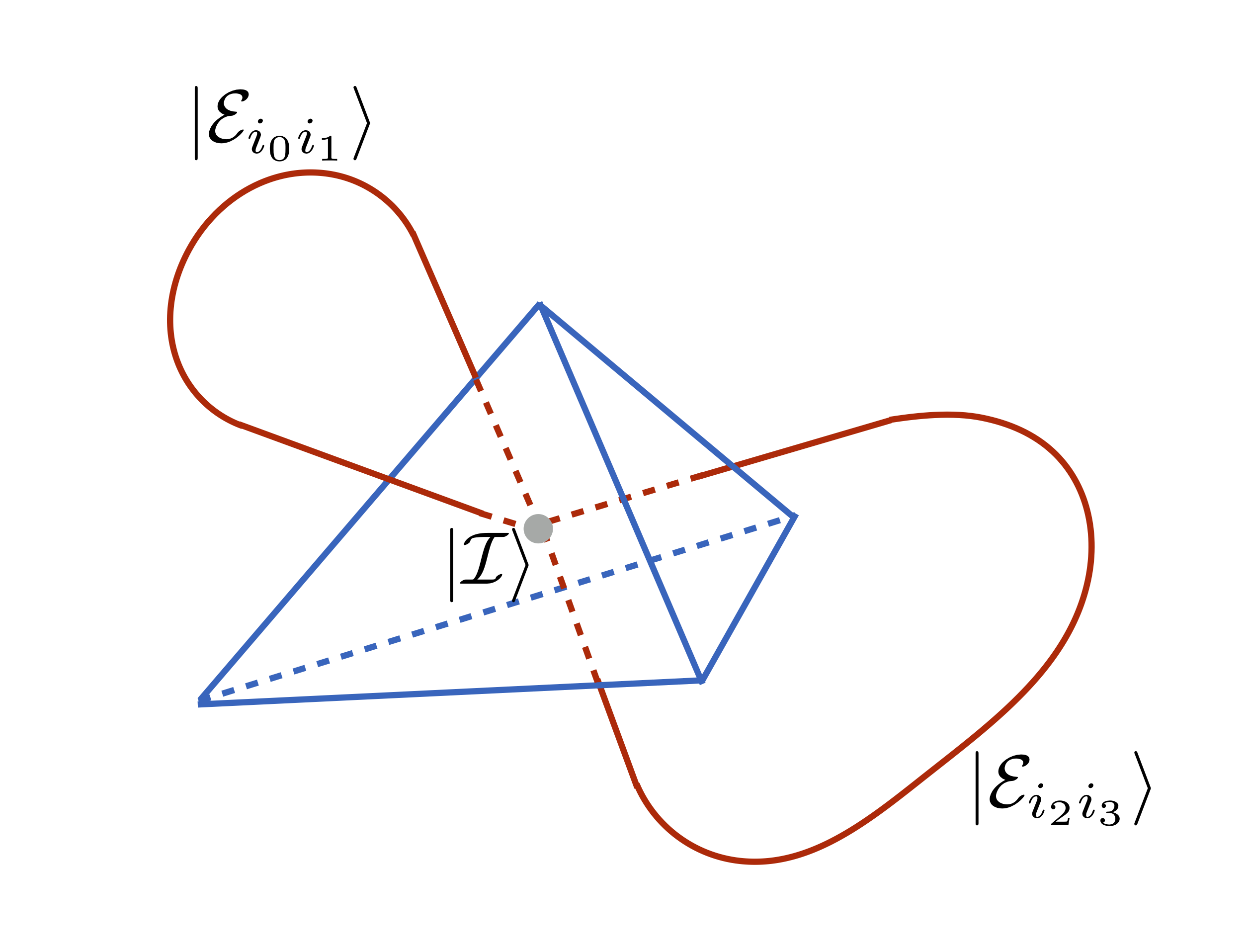}
	\caption{Monopole spin network and the corresponding pairing of the faces of the dual tetrahedron.}
	\label{MonopoleState}
\end{figure}

The three states are associated with connecting the faces of the dual tetrahedra as 
represented in Fig. \ref{MonopoleState}.

In the considered case the states are satisfying the Gauss constraint, therefore:
\begin{equation}
	\hat{P}_{\text{G}}|\mathcal{E}_{ijkl}\rangle=|\mathcal{E}_{ijkl}\rangle.
\end{equation}

The $U$ and $V$ matrices used in the quantum circuits are:
\begin{align}
U_{0123} &= \left(
\begin{array}{cc}
0 & 1 \\
-1 & 0
\end{array}	
\right) , V_{0123} = \frac{1}{\sqrt{2}}\left(
\begin{array}{cc}
-1 & -1 \\
1 & -1
\end{array}	
\right), \\
U_{0213}&=\frac{1}{\sqrt{2}}\left(
\begin{array}{cc}
1 & 1 \\
-1 & 1
\end{array}	
\right), 
V_{0213}=\left(
\begin{array}{cc}
0 & -1 \\
1 & 0
\end{array}	
\right), \\
U_{0312}&=\frac{1}{\sqrt{2}}\left(
\begin{array}{cc}
1 & 1 \\
-1 & 1
\end{array}	
\right),  V_{0312}=\left(
\begin{array}{cc}
-1 & 0 \\
0 & -1
\end{array}	
\right).
\end{align}

\begin{figure}[ht!]
	\leavevmode
	\centering
	\includegraphics[scale=1]{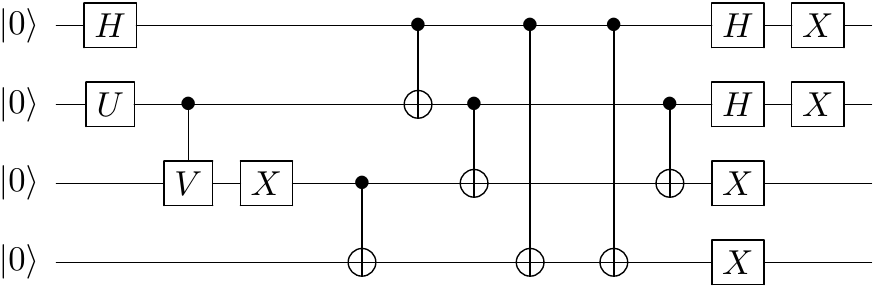}
	\caption{A quantum circuit for the transition amplitude $\langle\mathcal{I}|\mathcal{E}_{0312}\rangle$. Before reduction.}
	\label{MonopoleCircuit1}
\end{figure}

In Fig. \ref{MonopoleCircuit1} a quantum circuit associated with the amplitude $\langle\mathcal{I}|\mathcal{E}_{0312}\rangle$ is presented. 

\begin{figure}[ht!]
	\leavevmode
	\centering
	\includegraphics[scale=1]{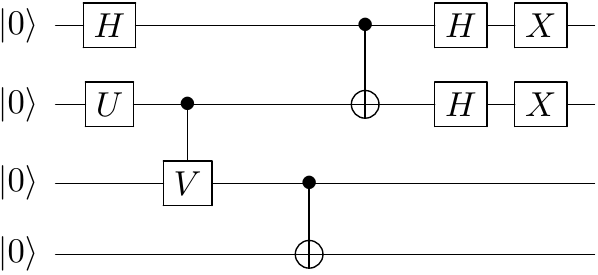}
	\caption{A quantum circuit for the transition amplitude $\langle\mathcal{I}|\mathcal{E}_{0312}\rangle$. After reduction.}
	\label{MonopoleCircuit2}
\end{figure}

The circuit can be further reduced to the form show in Fig. \ref{MonopoleCircuit2}.

Results of determination of probabilities $|\langle \mathcal{E}_{0312}|0_s\rangle|^2$ and 
$|\langle \mathcal{E}_{0312}|1_s\rangle|^2$ on Melbourne and Yorktown quantum computers are 
collected in Table \ref{MonopolTable}. In both cases the reduced circuit shown in Fig \ref{MonopoleCircuit2}
was ued.

\begin{table}
	\[
	\begin{array}{|c|c|c|c|}
	\hline
	\text{Amplitude} & \text{Theory}  & \text{Melbourne} & \text{Yorktown}\\
	\hline
	|\langle 0_s|\mathcal{E}_{0312}\rangle|^2 & 0.25 & 0.23\pm0.01 & 0.22\pm0.01\\
	|\langle 1_s|\mathcal{E}_{0312}\rangle|^2 & 0.75 & 0.72\pm0.01 & 0.67\pm0.01\\
	\hline
	\end{array}
	\]
	\caption{Results of simulations for the monopole spin network.}
	\label{MonopolTable}
\end{table}

In the simulations, a sequence of 10 computational rounds each containing 1024 shots 
was performed for every of the investigated states. The modulus squares of the amplitudes
were determined using the method introduced in Sec. \ref{Sec:Amplitudes}. In the considered
case, satisfactory agreement between the outcomes of measurement and the theoretical 
predictions are found, with slightly better results obtained with the use of the Melbourne
quantum processor. 

\section{Dipole Spin Network} \label{Sec:Dipole}

\begin{figure}[ht!]
	\centering
	\includegraphics[scale=0.25]{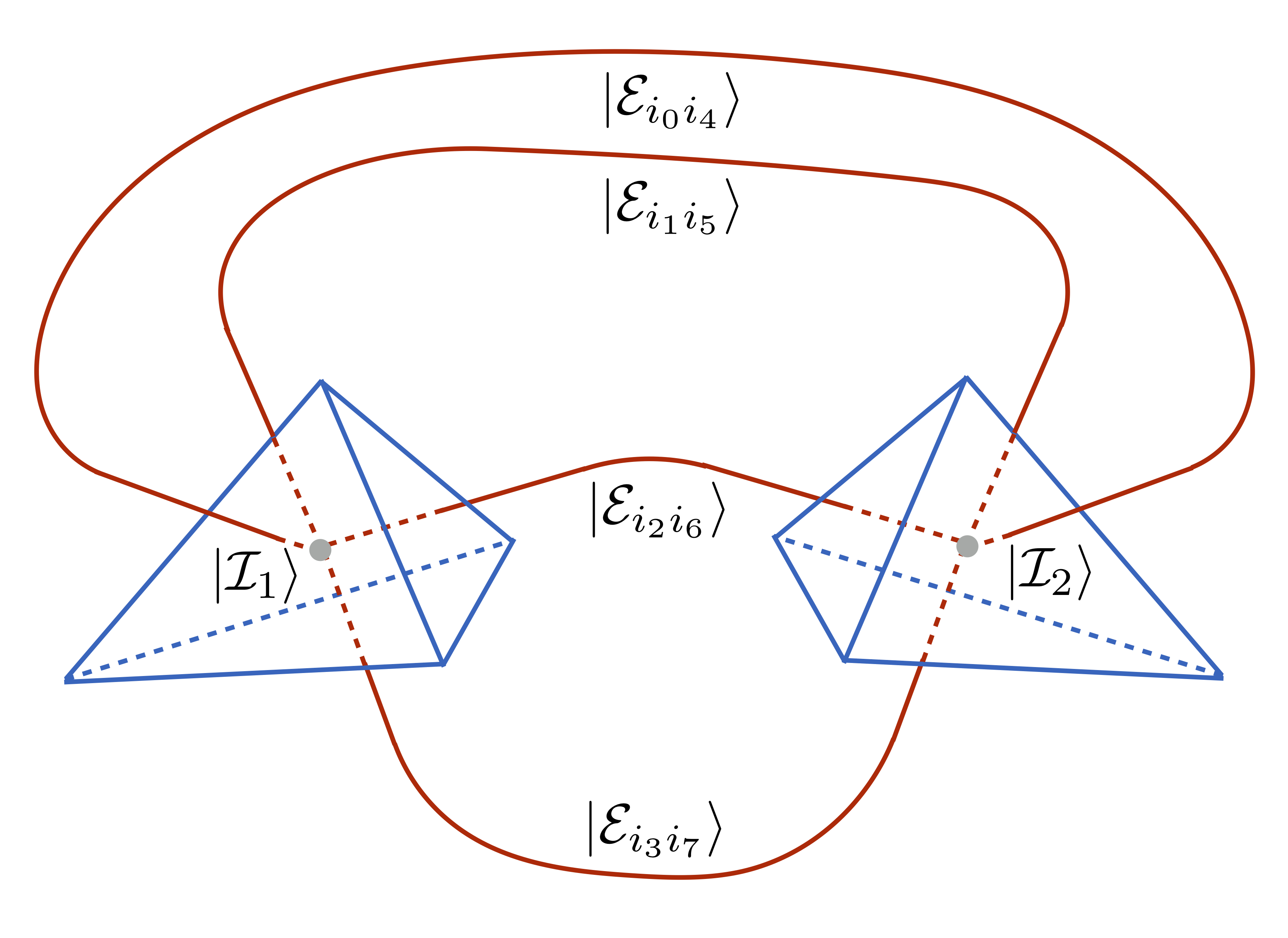}
	\caption{Dipole spin network and the corresponding pairing of the faces of the dual tetrahedra.}
	\label{DipoleState}
\end{figure}

In the geometric picture, dipole spin network is obtained by considering two 
tetrahedra glued together face by face, as depicted in Fig. \ref{DipoleState}.

Because, there are numerous possible permutations of the connections, there are 
various possible states of the maximally entangled states associated with the dipol diagram. 
The possible 24 configurations of connection and the corresponding states are summarized 
in Table \ref{DipolConnections}.
\begin{table}
	\[
	\begin{array}{|c|c|c|c|c|}
	\hline
	\text{Connections} & |0_s0_s\rangle & |0_s1_s\rangle & |1_s0_s\rangle & |1_s1_s\rangle \\
	\hline
	\{0, 4\}, \{1, 5\}, \{2, 6\}, \{3, 7\} & \multirow{4}{*}{$\frac{1}{4}$} & \multirow{4}{*}{$0$} & \multirow{4}{*}{$0$} & \multirow{4}{*}{$\frac{1}{4}$} \\
	\{0, 5\}, \{1, 4\}, \{2, 7\}, \{3, 6\} & & & & \\
	\{0, 6\}, \{1, 7\}, \{2, 4\}, \{3, 5\} & & & & \\
	\{0, 7\}, \{1, 6\}, \{2, 5\}, \{3, 4\} & & & & \\
	\hline
	\{0, 4\}, \{1, 5\}, \{2, 7\}, \{3, 6\} & \multirow{4}{*}{$-\frac{1}{4}$} & \multirow{4}{*}{$0$} & \multirow{4}{*}{$0$} & \multirow{4	}{*}{$\frac{1}{4}$} \\
	\{0, 5\}, \{1, 4\}, \{2, 6\}, \{3, 7\} & & & & \\
	\{0, 6\}, \{1, 7\}, \{2, 5\}, \{3, 4\} & & & & \\
	\{0, 7\}, \{1, 6\}, \{2, 4\}, \{3, 5\} & & & & \\
	\hline
	\{0, 4\}, \{1, 6\}, \{2, 5\}, \{3, 7\} & \multirow{4}{*}{$\frac{1}{8}$} & \multirow{4}{*}{$\frac{\sqrt{3}}{8}$} & \multirow{4}{*}{$\frac{\sqrt{3}}{8}$} & \multirow{4}{*}{$-\frac{1}{8}$} \\
	\{0, 5\}, \{1, 7\}, \{2, 4\}, \{3, 6\} & & & & \\
	\{0, 6\}, \{1, 4\}, \{2, 7\}, \{3, 5\} & & & & \\
	\{0, 7\}, \{1, 5\}, \{2, 6\}, \{3, 4\} & & & & \\
	\hline
	\{0, 4\}, \{1, 6\}, \{2, 7\}, \{3, 5\} & \multirow{4}{*}{$-\frac{1}{8}$} & \multirow{4}{*}{$-\frac{\sqrt{3}}{8}$} & \multirow{4}{*}{$\frac{\sqrt{3}}{8}$} & \multirow{4}{*}{$-\frac{1}{8}$} \\
	\{0, 5\}, \{1, 7\}, \{2, 6\}, \{3, 4\} & & & & \\
	\{0, 6\}, \{1, 4\}, \{2, 5\}, \{7, 3\}  & & & & \\
	\{0, 7\}, \{1, 5\}, \{2, 4\}, \{6, 3\}  & & & & \\
	\hline
	\{0, 4\}, \{1, 7\}, \{2, 5\}, \{3, 6\} & \multirow{4}{*}{$-\frac{1}{8}$} & \multirow{4}{*}{$\frac{\sqrt{3}}{8}$} & \multirow{4}{*}{$-\frac{\sqrt{3}}{8}$} & \multirow{4}{*}{$-\frac{1}{8}$} \\
	\{0, 5\}, \{1, 6\}, \{2, 4\}, \{3, 7\} & & & & \\
	\{0, 6\}, \{1, 5\}, \{2, 7\}, \{3, 4\} & & & & \\
	\{0, 7\}, \{1, 4\}, \{2, 6\},\{3, 5\} & & & & \\
	\hline
	\{0, 4\}, \{1, 7\}, \{2, 6\}, \{3, 5\} & \multirow{4}{*}{$\frac{1}{8}$} & \multirow{4}{*}{$-\frac{\sqrt{3}}{8}$} & \multirow{4}{*}{$-\frac{\sqrt{3}}{8}$} & \multirow{4}{*}{$-\frac{1}{8}$} \\
	\{0, 5\}, \{1, 6\}, \{2, 7\}, \{3, 4\} & & & & \\
	\{0, 6\}, \{1, 5\}, \{2, 4\}, \{3, 7\} & & & & \\
	\{0, 7\}, \{1, 4\}, \{2, 5\}, \{3, 6\} & & & & \\
	\hline
	\end{array}
	\]
	\caption{Amplitudes of the projected states for the 24 combinations of connections for the dipol diagram.}
	\label{DipolConnections}
\end{table}

As an example, we will consider the following state:
\begin{equation}
	|\mathcal{E}_{04152637}\rangle =|\mathcal{E}_{04}\rangle|\mathcal{E}_{15}\rangle|\mathcal{E}_{26}\rangle|\mathcal{E}_{37}\rangle,
\end{equation}
which corresponds to the connections $\{\{0, 4\},\{1, 5\},\{2, 6\},\{3, 7\}\}$. Projecting the state onto 
the spin network basis (imposing the Gauss constraint) gives, 
\begin{equation}
	\hat{P}_{\text{G}}|\mathcal{E}_{04152637}\rangle =\frac{1}{4}\left(|0_s0_s\rangle+|1_s1_s\rangle\right)
\end{equation}
such that in consequence, we have the two non-vanishing amplitudes:
\begin{align}
& \langle 0_s0_s  |\mathcal{E}_{04152637}\rangle = \langle 1_s1_s  |\mathcal{E}_{04152637}\rangle = \frac{1}{4},  \\
& \langle 0_s1_s  |\mathcal{E}_{04152637}\rangle = \langle 1_s0_s  |\mathcal{E}_{04152637}\rangle =0. 
\end{align}

\begin{figure*}[ht!]
	\leavevmode
	\centering
	\includegraphics[scale=1]{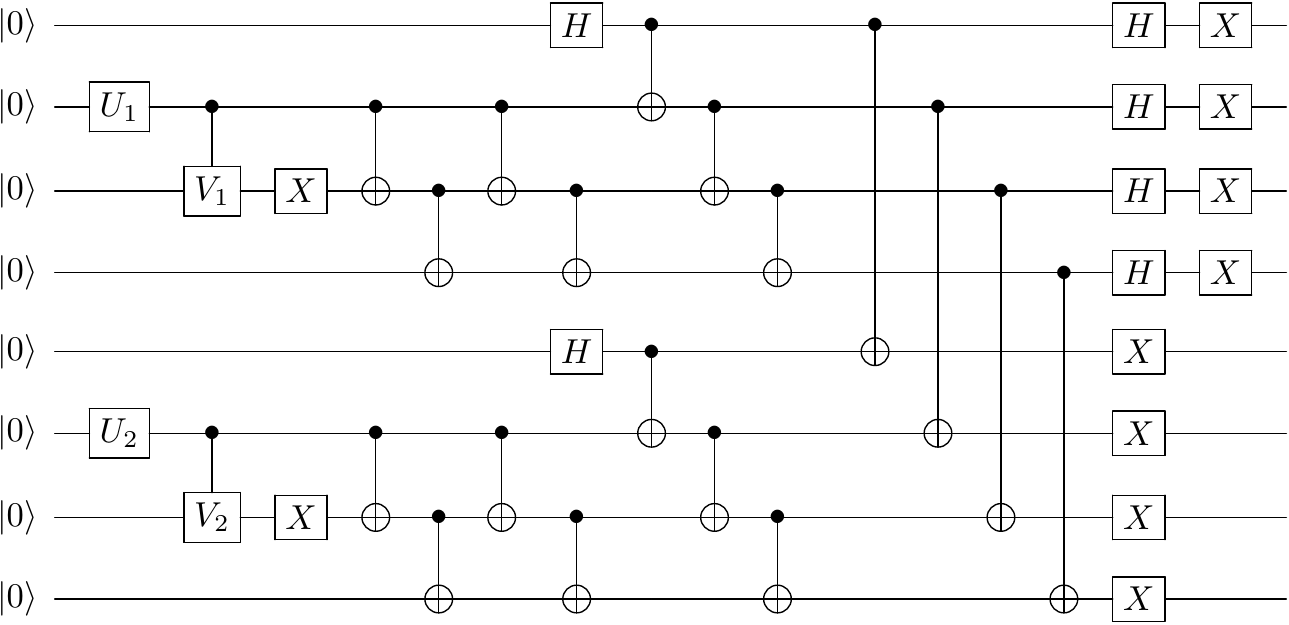}
	\caption{A quantum circuit for transition amplitude $\langle\mathcal{I}_1\mathcal{I}_2|\mathcal{E}_{04152637}\rangle$.}
	\label{DipolCircuit1}
\end{figure*}

\begin{figure}[ht!]
	\leavevmode
	\centering
	\includegraphics[scale=1]{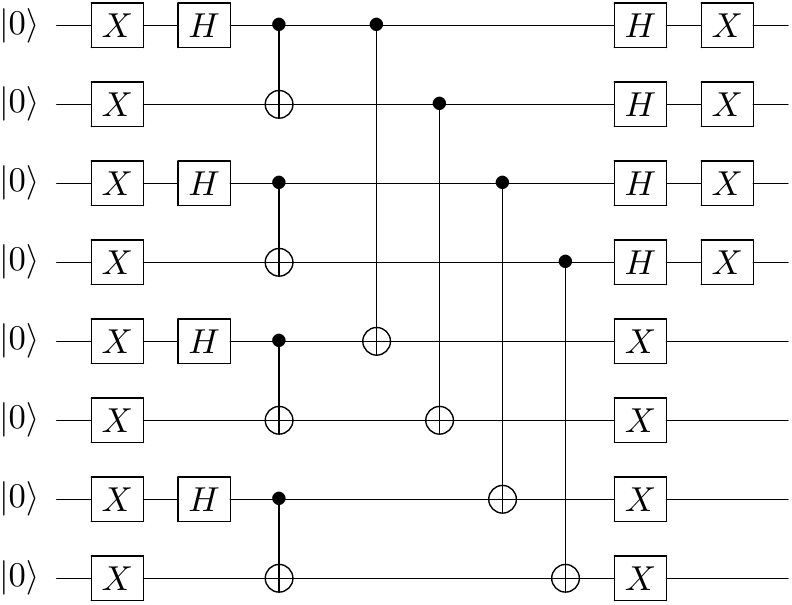}
	\caption{Simplified version of a quantum circuit for transition amplitude $\langle0_s0_s|\mathcal{E}_{04152637}\rangle$.}
		\label{DipolCircuit2}
\end{figure}

From the viewpoint of quantum computing, the amplitudes can be determined by evaluating
the quantum circuit presented in Fig. \ref{DipolCircuit1}.
For the special case of the states of the interwiners $|\mathcal{I}_1\mathcal{I}_2\rangle = |0_s 0_s \rangle$ the 
quantum circuit can be simplified to the form presented in Fig. \ref{DipolCircuit2}, where representation 
of the state $|0_s \rangle$ by the circuit (\ref{0sStateOld}) has been used. 
 
The circuits can be directly embedded into the architecture of the Melbourne quantum processors, 
shown in Fig. \ref{Melbourne}. Results of our simulations are collected in Table \ref{DipoleTableOutcomes}.
\begin{table}
	\[
	\begin{array}{|c|c|c|c|}
	\hline
	\text{Amplitude} & \text{Theory} & \text{Melbourne} & \text{Melbourne - S}\\
	\hline
	|\langle 0_s0_s|\mathcal{E}_{04152637}\rangle|^2 & 0.0625 & 0.008\pm0.002 & 0.003\pm0.001\\
	|\langle 0_s1_s|\mathcal{E}_{04152637}\rangle|^2 & 0 & 0.003\pm0.002 & -\\
	|\langle 1_s0_s|\mathcal{E}_{04152637}\rangle|^2 & 0 & 0.009\pm0.002 & -\\
	|\langle 1_s1_s|\mathcal{E}_{04152637}\rangle|^2 & 0.0625 & 0.008\pm0.003 & -\\
	\hline
	\end{array}
	\]
	\caption{Results of simulations for the dipole network on the 15 qubit Melbourne 
	quantum computer.}
	\label{DipoleTableOutcomes}
\end{table}

As previously, a sequence of 10 computational rounds each containing 1024 shots 
was performed for every of the investigated states. Eight out of fifteen qubits of the 
Melbourne processor have been used in the computations. The third column of Table 
\ref{DipoleTableOutcomes} contains results of simulations based on the circuit shown 
in Fig.  \ref{DipolCircuit1}) whereas the fourth column presents result obtained using 
the circuit shown in Fig. \ref{DipolCircuit2}. The obtained results differ cardinally
from the theoretical predictions. The reason for this is most probably significant depth 
of the considered quantum circuits and accumulations of errors.  In order to better 
understand this issue in the employed quantum chip a set of test have been performed.
Results of the tests are collected in Appendix B. One can find that significant accumulation
of errors is present even for simple low-depth circuits. This indicates, that going 
beyond the case of a single node (with the use of currently available quantum processors)
cannot be done successfully without adopting quantum error correction codes.  

\section{Summary} \label{Sec:Summary}

Loop quantum gravity and related approaches to gravity, such as Group Field Theories,  
provide picture of spacetime as a many-body quantum system \cite{Oriti:2017twl}. The 
degrees of freedom are associated with the quanta of volume (``atoms of space") related 
to the nodes of a spin network. This viewpoint opens an interesting possibility to 
employ many-body quantum physics methods designed to explore complex collective 
properties of composite systems. Especially promising paths to include: tensor 
networks methods and quantum simulations. 

In this article, the second method has been discussed, following the ideas developed in 
Refs. \cite{Li:2017gvt,Mielczarek:2018ttq,Mielczarek:2018jsh}. Primarily, our focus was 
on construction of a quantum circuit for a general intertwiner qubit state. Such circuit
has been introduced and shown to utilize four logical qubits, without involving quantum 
error correction codes. The presented circuit is a generalization of the circuit for the 
basis state $|0_s\rangle$ explored in Ref. \cite{Mielczarek:2018jsh}. Based on the circuit, 
exemplary intertwiner qubits states were simulated on both 5-qubit (Yorktown) and 
15-qubit (Melbourne) IBM superconducting quantum processors. It has been shown that 
for the case of the 5-qubit machine, fidelities of the obtained states reach the level of 
$F\approx 90$ \%. On the other hand, while the total number of qubits of the processor 
is increased to 15, the fidelity of the states drops down to $F\approx 85$ \%, even if the 
number of utilized logical qubits remains to be four. This is a first sign of the fact that 
it is much more difficult to keep quantum coherence of bigger quantum systems. 
Further, even more drastic, consequences of this fact have beed observed while 
transition amplitudes between simple spin network states were studied. 

For this purpose, a class of maximally entangled spin network states, analogous
to the PEPS tensor networks, has been introduced. The states have been introduced 
by considering maximally entangled states between source and target Hilbert spaces 
of holonomies, corresponding to links of the spin network. Such possibility is supported 
by recent results presented in Ref. \cite{Mielczarek:2019srn}. Furthermore, the state 
of maximally entangled links has to be projected onto the surface of Gauss constraint 
in order to get well defined superposition of spin network states. With the use of such 
appropriately projected state, exemplary transition amplitudes for a monopole and 
dipole spin networks have been considered. 

The monopole spin network amplitudes required only four logical qubits, and runs 
of the associated quantum circuit on a superconducting 5-qubit chip lead to  
good agreement with theoretical predictions. On the other hand, the dipole spin 
network involves 8 logical qubits, and the associated quantum circuit was transpilated 
to the form compatible with topology of the available 15 qubit IBM quantum processor. 
However, because of significant errors, running of the circuit on the quantum 
computer did not lead to reasonable results. Therefore, for the moment, quantum 
simulations of the dipole spin networks are still challenging. This concerns 
the considered publicly available IBM superconducting quantum process, which 
has been used. However, the current hight activity in the quantum computing technologies 
prognosis that both the dipole and more complex spin networks will be possible 
simulate successfully in the coming years.  

\section{Acknowledgements}

Authors are supported from the Sonata Bis Grant DEC-2017/26/E/ST2/00763 of the 
National Science Centre Poland. Furthermore,  this publication was made possible 
through the support of the ID\# 61466 grant from the John Templeton Foundation, 
as part of the ``The Quantum Information Structure of Spacetime (QISS)'' Project 
(qiss.fr). The opinions expressed in this publication are those of the authors 
and do not necessarily reflect the views of the John Templeton Foundation. Authors 
would also like to thank to Daniel Nagaj for discussion on the quantum circuits at the 
initial stage of the preparation of the article. 

\section*{Appendix A} \label{Sec:AppendixA}

The appendix summarizes numerical data obtained from evaluation of the 
quantum circuits for the intertwiner qubit states on IBM superconducting quantum 
computers. For each of the considered case, 10 computational rounds have been 
performed each of 1024 shots (evaluation of quantum circuit and performing 
measurement). Both averages and standard deviations have been determined 
based on the 10 computational rounds.  

In Table \ref{FullTableYorktown} the results of quantum simulations on the 5-qubit 
(Yorktown) IBM quantum computer are collected.   

\begin{table*}
	\[
	\begin{array}{|c|c|c|c|c|c|c|}
	\hline 
	\text{State}&|0_s\rangle&|1_s\rangle&|+\rangle &|-\rangle&|\circlearrowright\rangle&|\circlearrowleft\rangle\\ 
	\hline
	|0000\rangle & 0.007\pm0.001 & 0.021\pm0.003 & 0.016\pm0.005 & 0.008\pm0.003 & 0.01\pm0.003 & 0.009\pm0.003 \\
	|0001\rangle & 0.020\pm0.002 & 0.014\pm0.006 & 0.015\pm0.004 & 0.016\pm0.004 & 0.015\pm0.004 & 0.02\pm0.005 \\
	|0010\rangle & 0.024\pm0.006 & 0.024\pm0.005 & 0.025\pm0.006 & 0.028\pm0.007 & 0.027\pm0.005 & 0.022\pm0.004 \\
	|0011\rangle & 0.008\pm0.003 & 0.305\pm0.015 & 0.172\pm0.008 & 0.128\pm0.015 & 0.151\pm0.014 & 0.142\pm0.010 \\
	|0100\rangle & 0.023\pm0.007 & 0.016\pm0.004 & 0.012\pm0.003 & 0.025\pm0.004 & 0.017\pm0.005 & 0.022\pm0.006 \\
	|0101\rangle & 0.183\pm0.011 & 0.066\pm0.007 & 0.03\pm0.004 & 0.243\pm0.013 & 0.123\pm0.012 & 0.150\pm0.006 \\
	|0110\rangle & 0.175\pm0.012 & 0.048\pm0.007 & 0.193\pm0.010 & 0.026\pm0.003 & 0.126\pm0.01 & 0.099\pm0.012 \\
	|0111\rangle & 0.017\pm0.004 & 0.009\pm0.003 & 0.021\pm0.005 & 0.007\pm0.003 & 0.013\pm0.005 & 0.011\pm0.005 \\
	|1000\rangle & 0.025\pm0.005 & 0.021\pm0.004 & 0.030\pm0.006 & 0.018\pm0.005 & 0.025\pm0.005 & 0.026\pm0.008 \\
	|1001\rangle & 0.206\pm0.013 & 0.069\pm0.010 & 0.200\pm0.016 & 0.059\pm0.005 & 0.119\pm0.007 & 0.158\pm0.008 \\
	|1010\rangle & 0.262\pm0.012 & 0.077\pm0.007 & 0.087\pm0.005 & 0.277\pm0.016 & 0.190\pm0.009 & 0.160\pm0.009 \\
	|1011\rangle & 0.022\pm0.006 & 0.014\pm0.002 & 0.012\pm0.004 & 0.024\pm0.004 & 0.017\pm0.004 & 0.021\pm0.004 \\
	|1100\rangle & 0.007\pm0.003 & 0.278\pm0.011 & 0.151\pm0.01 & 0.119\pm0.007 & 0.141\pm0.011 & 0.138\pm0.009 \\
	|1101\rangle & 0.007\pm0.003 & 0.010\pm0.004 & 0.011\pm0.003 & 0.010\pm0.003 & 0.009\pm0.004 & 0.007\pm0.003 \\
	|1110\rangle & 0.008\pm0.002 & 0.008\pm0.003 & 0.010\pm0.003 & 0.005\pm0.001 & 0.007\pm0.003 & 0.006\pm0.002 \\
	|1111\rangle & 0.006\pm0.002 & 0.020\pm0.003 & 0.015\pm0.003 & 0.007\pm0.002 & 0.011\pm0.005 & 0.011\pm0.002 \\
	\hline
	\end{array}
	\]
	\caption{Experimental results for the states generated on 5-qubit (Yorktown) IBM quantum computer.} 
	\label{FullTableYorktown}
\end{table*}

In Table \ref{FullTableMelbourne} the results of quantum simulations on the 15-qubit (Melbourne) IBM quantum 
computer are collected.   

\begin{table*}
	\[
	\begin{array}{|c|c|c|c|c|c|c|}
	\hline 
	\text{State}&|0_s\rangle&|1_s\rangle&|+\rangle &|-\rangle&|\circlearrowright\rangle&|\circlearrowleft\rangle\\ 
	\hline
	|0000\rangle & 0.04\pm0.005 & 0.019\pm0.003 & 0.043\pm0.005 & 0.013\pm0.003 & 0.026\pm0.005 & 0.024\pm0.006 \\
	|0001\rangle & 0.033\pm0.005 & 0.053\pm0.008 & 0.05\pm0.007 & 0.035\pm0.008 & 0.041\pm0.006 & 0.05\pm0.006 \\
	|0010\rangle & 0.033\pm0.006 & 0.033\pm0.006 & 0.026\pm0.005 & 0.042\pm0.006 & 0.036\pm0.007 & 0.034\pm0.004 \\
	|0011\rangle & 0.025\pm0.005 & 0.239\pm0.012 & 0.109\pm0.008 & 0.119\pm0.011 & 0.096\pm0.009 & 0.16\pm0.008 \\
	|0100\rangle & 0.034\pm0.005 & 0.043\pm0.007 & 0.032\pm0.005 & 0.039\pm0.006 & 0.034\pm0.006 & 0.035\pm0.007 \\
	|0101\rangle & 0.128\pm0.011 & 0.105\pm0.009 & 0.033\pm0.005 & 0.225\pm0.014 & 0.158\pm0.014 & 0.085\pm0.011 \\
	|0110\rangle & 0.246\pm0.015 & 0.063\pm0.006 & 0.244\pm0.011 & 0.055\pm0.007 & 0.156\pm0.011 & 0.158\pm0.008 \\
	|0111\rangle & 0.028\pm0.005 & 0.014\pm0.002 & 0.018\pm0.006 & 0.026\pm0.003 & 0.017\pm0.004 & 0.02\pm0.004 \\
	|1000\rangle & 0.029\pm0.006 & 0.02\pm0.005 & 0.025\pm0.005 & 0.026\pm0.003 & 0.026\pm0.004 & 0.026\pm0.006 \\
	|1001\rangle & 0.166\pm0.019 & 0.065\pm0.008 & 0.207\pm0.015 & 0.037\pm0.007 & 0.129\pm0.009 & 0.126\pm0.007 \\
	|1010\rangle & 0.135\pm0.013 & 0.065\pm0.009 & 0.035\pm0.007 & 0.217\pm0.015 & 0.148\pm0.008 & 0.082\pm0.007 \\
	|1011\rangle & 0.011\pm0.004 & 0.024\pm0.006 & 0.014\pm0.004 & 0.022\pm0.005 & 0.016\pm0.004 & 0.017\pm0.003 \\
	|1100\rangle & 0.024\pm0.005 & 0.204\pm0.016 & 0.094\pm0.008 & 0.1\pm0.007 & 0.064\pm0.011 & 0.126\pm0.009 \\
	|1101\rangle & 0.014\pm0.004 & 0.018\pm0.004 & 0.012\pm0.006 & 0.021\pm0.005 & 0.016\pm0.005 & 0.015\pm0.003 \\
	|1110\rangle & 0.023\pm0.007 & 0.018\pm0.003 & 0.027\pm0.008 & 0.012\pm0.003 & 0.019\pm0.005 & 0.022\pm0.002 \\
	|1111\rangle & 0.031\pm0.006 & 0.016\pm0.005 & 0.03\pm0.004 & 0.012\pm0.004 & 0.017\pm0.003 & 0.02\pm0.005 \\
	\hline
	\end{array}
	\]
	\caption{Experimental results for the states generated on 15-qubit (Melbourne) IBM quantum computer.}
	\label{FullTableMelbourne}
\end{table*}

For comparison, in Table \ref{FullTableTheory} theoretical values of the probabilities of the basis states for the 
states under consideration are shown. 

\begin{table}
	\[
	\begin{array}{|c|c|c|c|c|c|c|}
	\hline 
	\text{State}&|0_s\rangle&|1_s\rangle&|+\rangle &|-\rangle&|\circlearrowright\rangle&|\circlearrowleft\rangle\\ 
	\hline
	|0000\rangle & 0.0 & 0.0 & 0.0 & 0.0 & 0.0 & 0.0 \\
	|0001\rangle & 0.0 & 0.0 & 0.0 & 0.0 & 0.0 & 0.0 \\
	|0010\rangle & 0.0 & 0.0 & 0.0 & 0.0 & 0.0 & 0.0 \\
	|0011\rangle & 0.0 & 0.333 & 0.167 & 0.167 & 0.167 & 0.167 \\
	|0100\rangle & 0.0 & 0.0 & 0.0 & 0.0 & 0.0 & 0.0 \\
	|0101\rangle & 0.25 & 0.083 & 0.022 & 0.311 & 0.167 & 0.167 \\
	|0110\rangle & 0.25 & 0.083 & 0.311 & 0.022 & 0.167 & 0.167 \\
	|0111\rangle & 0.0 & 0.0 & 0.0 & 0.0 & 0.0 & 0.0 \\
	|1000\rangle & 0.0 & 0.0 & 0.0 & 0.0 & 0.0 & 0.0 \\
	|1001\rangle & 0.25 & 0.083 & 0.311 & 0.022 & 0.167 & 0.167 \\
	|1010\rangle & 0.25 & 0.083 & 0.022 & 0.311 & 0.167 & 0.167 \\
	|1011\rangle & 0.0 & 0.0 & 0.0 & 0.0 & 0.0 & 0.0 \\
	|1100\rangle & 0.0 & 0.333 & 0.167 & 0.167 & 0.167 & 0.167 \\
	|1101\rangle & 0.0 & 0.0 & 0.0 & 0.0 & 0.0 & 0.0 \\
	|1110\rangle & 0.0 & 0.0 & 0.0 & 0.0 & 0.0 & 0.0 \\
	|1111\rangle & 0.0 & 0.0 & 0.0 & 0.0 & 0.0 & 0.0 \\
	\hline
	\end{array}
	\]
	\caption{Theoretical probabilities for the states under consideration.}
	\label{FullTableTheory}
\end{table}

\section*{Appendix B} \label{Sec:AppendixB}

The appendix summarizes tests performed on the 15 qubits IBM quantum processor Melbourne.
The following four tests have been performed:  
\begin{enumerate}
	\item Measurements on $n$ qubits without any quantum gates applied (the $\otimes_{i=1}^{15} |0 \rangle$ state).
	\item Applying NOT gates ($\hat{X}$) and measurement on $n$ qubits. 
	\item Applying NOT gates on all 15 qubits and performing measurement on $n$ first qubits. 
	\item Applying NOT gates on $n$ qubits and performing measurement on all 15 qubits.
\end{enumerate}

\begin{figure*}
	\includegraphics[scale=0.4]{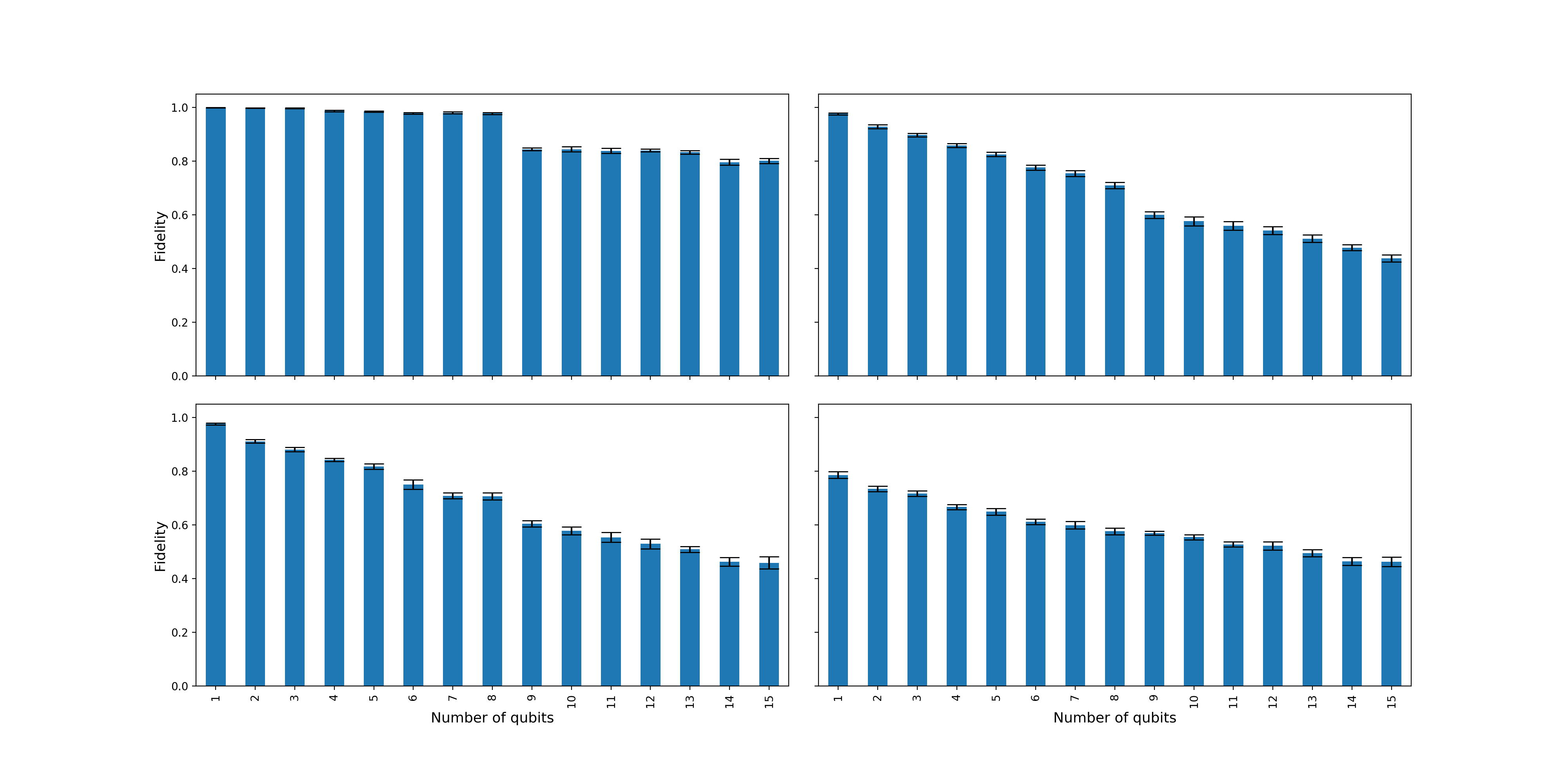}
	\caption{Fidelities for the four types of tests performed on the 15-qubit Melbourne quantum 
	computer. From top left: measures on $n$ qubits for the empty quantum register, NOT gates 
	and measures on $n$ qubits, NOT gates on all 15 qubits and measurement on $n$ 
	qubits, NOT gates on $n$ qubits and measuremen on all 15 qubits.}
	\label{TestsFidelities}
\end{figure*}

In Fig. \ref{TestsFidelities} fidelities for the states obtained for the four test are presented. The presented 
data are collected in Table \ref{FidelitiesTests}.  

\begin{table}
	\[
	\begin{array}{|c|c|c|c|c|}
	\hline
	n&1^{st}\ \text{test}&2^{nd}\ \text{test} & 3^{rd}\ \text{test} & 4^{th}\ \text{test}\\
	\hline
	1&1.000\pm0.000&0.976\pm0.003&0.976\pm0.003&0.786\pm0.012\\
	2&0.998\pm0.001&0.928\pm0.007&0.911\pm0.006&0.734\pm0.010\\
	3&0.997\pm0.002&0.897\pm0.007&0.881\pm0.008&0.717\pm0.011\\
	4&0.987\pm0.003&0.859\pm0.007&0.843\pm0.006&0.667\pm0.009\\
	5&0.985\pm0.002&0.825\pm0.008&0.818\pm0.010&0.649\pm0.012\\
	6&0.979\pm0.003&0.776\pm0.009&0.751\pm0.018&0.611\pm0.010\\
	7&0.980\pm0.004&0.755\pm0.011&0.709\pm0.011&0.599\pm0.014\\
	8&0.978\pm0.004&0.710\pm0.012&0.707\pm0.013&0.576\pm0.013\\
	9&0.844\pm0.006&0.600\pm0.013&0.604\pm0.012&0.569\pm0.008\\
	10&0.845\pm0.009&0.576\pm0.017&0.579\pm0.015&0.554\pm0.009\\
	11&0.839\pm0.010&0.560\pm0.016&0.554\pm0.018&0.528\pm0.009\\
	12&0.840\pm0.005&0.542\pm0.015&0.529\pm0.018&0.522\pm0.015\\
	13&0.834\pm0.007&0.512\pm0.014&0.509\pm0.011&0.495\pm0.013\\
	14&0.797\pm0.011&0.478\pm0.011&0.463\pm0.017&0.464\pm0.015\\
	15&0.801\pm0.010&0.438\pm0.014&0.459\pm0.023&0.463\pm0.017\\
	\hline
	\end{array}
	\]
	\caption{Fidelities of states for the four tests performed on the 15 qubit IBM 
	quantum computer (Melbourne).}
	\label{FidelitiesTests}
\end{table}

 \end{document}